\documentclass[amsmath, a4paper, pra, twocolumn, showpacs]{revtex4}

\usepackage{amssymb}
\usepackage{amsmath}
\usepackage{epsfig}
\usepackage{color}
\usepackage{graphics, graphicx}
\usepackage{bbold}
\usepackage{psfrag}
\usepackage{mathcomp}

\begin{document}

\author{Michiel Snoek}
%\author{Masudul Haque}
%\author{S. Vandoren}
\author{H. T. C. Stoof}
\affiliation{Institute for Theoretical Physics, Utrecht University, Leuvenlaan
4, 3584 CE Utrecht, The
Netherlands}
%\date{\today}

\title{Theory of vortex-lattice melting in a one-dimensional optical lattice}

\begin{abstract}
We investigate quantum and temperature fluctuations of a vortex lattice in a
one-dimensional optical lattice. We discuss in particular the Bloch bands of the
Tkachenko modes and calculate the correlation function of the vortex positions
along the
direction of the optical lattice.
Because of the small
number of particles in the pancake Bose-Einstein condensates at every site of
the optical lattice, finite-size effects become very
important. Moreover, the 
fluctuations in the vortex positions are inhomogeneous due to 
the inhomogeneous density. As a result, the melting of the
lattice occurs from the outside inwards. 
However, tunneling between neighboring pancakes substantially reduces the
inhomogeneity as well as the size of the fluctuations. 
On the other hand, nonzero temperatures increase the size of the fluctuations
dramatically. We
calculate the
crossover temperature from
quantum melting to classical melting.
We also investigate melting in the presence of a quartic radial potential, where
a
liquid can form in the center instead of at the outer edge of the pancake
Bose-Einstein condensates.
\end{abstract}

\maketitle

\section{Introduction} 
Since the onset of experiments on Bose-Einstein condensates, vortices have
attracted a lot of attention. 
When the Bose-Einstein condensate is rotated faster than some critical rotation
frequency $\Omega_c$, a vortex appears in the gas \cite{Matthews99, Madison00,
Fetter01_JP}. Upon increasing the rotation frequency further, the number of
vortices
increases
\cite{Hodby01, Hodby02, Rosenbusch02, Leanhardt02, Leanhardt03}, and the
vortices order themselves in a hexagonal Abrikosov lattice \cite{Abo-Shaeer01,
Abo-Shaeer02, Engels02, Engels03, Coddington03, Schweikhard04, Bretin04}. 
If the rotation frequency is increased even further, the very rapidly rotating
ultracold bosonic gases have been
predicted to form 
highly-correlated quantum states \cite{Wilkin98, Cooper99, Wilkin00, Cooper01,
Ho02}.
In these states, the Bose-Einstein condensate has been completely depleted by
quantum fluctuations, and quantum liquids appear with excitations that can carry
fractional statistics. Some of these states have been identified with (bosonic)
fractional quantum Hall states \cite{Cooper01, Paredes02, Regnault03}, such as
the Laughlin state \cite{Laughlin83}, the Moore-Read state \cite{Moore91} and
various
Read-Rezayi states \cite{Read99, Rezayi05}.

In this article we study the quantum fluctuations of  vortices in a
one-dimensional optical lattice.
Optical lattices provide a powerful tool in manipulating a Bose-Einstein
condensate. By using a three-dimensional optical lattice, the Bose-Hubbard model
\cite{Jaksch98}
was experimentally realized  and the superfluid-Mott insulator
transition was observed \cite{Greiner02}. Moreover, a two-dimensional optical
lattice was used to create one-dimensional Bose-gases and to study the crossover
between a superfluid Bose gas and the ``fermionized'' Tonks gas \cite{Paredes04,
Kinoshita04}.
Theoretically, the
combination of a two-dimensional optical lattice and a single vortex was
predicted to give interesting effects around the superfluid-Mott insulator
transition \cite{Wu04}.
Optical lattices can also be used to
provide a pinning potential for vortices \cite{Reijnders04, Reijnders05, Pu05,
Bhat06, Burkov06}, and experiments on this topic are ongoing \cite{Cornell05}.
The physics of a single vortex line in a one-dimensional optical lattice is
recently extensively studied  
\cite{Martikainen03, Martikainen03B, Martikainen04, Martikainen04B,
Martikainen04C, Isoshima05}. 
By putting fermions in the vortex core, this system can even be used to create a
superstring in the laboratory \cite{Snoek05}.

\begin{figure}
\begin{center}
\vspace{-.5cm}
\includegraphics[scale=.25, angle=270, origin=c]{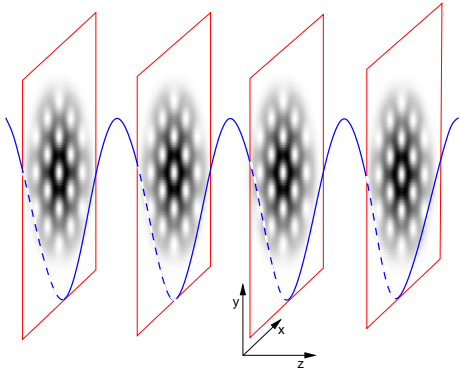}
\vspace{-.5cm}
\end{center}
\caption{(Color online) Setup in which the melting of the vortex lattice is
studied. 
An optical lattice along the $z$ direction divides the
condensate 
into pancake condensate which are coupled by tunneling processes.}
\label{cartoonl}
\end{figure}

A one-dimensional optical lattice divides the Bose-Einstein condensate into a
stack of
two-dimensional pancake condensates that are weakly coupled by tunneling as
schematically 
shown in Fig. \ref{cartoonl}. An important consequence of this setup is that the
modes of
the on-site two-dimensional vortex lattice form Bloch bands as a function of the
axial
momentum. In this paper we pay special attention to the Tkachenko modes
\cite{Tkachenko66},
which are the  lowest-lying modes of the vortex lattice.
Recently, these modes have been investigated for a single two-dimensional vortex
lattice both experimentally
\cite{Schweikhard04, Coddington03} and theoretically 
\cite{Baym03, Mizushima04, Baksmaty04, Baym04, Kim04, Cozzini04, Sonin05,
Sonin05B}.

The number of particles in a single
pancake is much smaller than in a Bose-Einstein condensate in a harmonic
trap, and hence the fluctuations are much 
larger. Therefore, this system is a promising candidate to reach the quantum
Hall regime.
The requirement for that is that  the
ratio $\nu=N/N_v$ of the number of atoms $N$ and the number of vortices $N_v$
is smaller than a critical value $\nu_c$. The ratio $\nu$ plays the
role of the filling factor and estimates for the
critical $\nu_c$ are typically around $8$ \cite{Cooper01, Sinova02}. 
However, observed filling factors are up till now always greater than 100, where
almost
perfect hexagonal lattices form and no sign of melting can be seen
\cite{Schweikhard04}. These experiments are carried out with Bose-Einstein
condensates consisting of typically $10^5$ particles, whereas the maximum number
of vortices observed is around $300$. Decreasing the number of particles results
in loss of signal, whereas the number of vortices is limited by the
rotation frequency that has to be smaller than the transverse trapping
frequency. Adding a quartic potential, which stabilizes the condensate also for
rotation frequencies higher than the transverse trap frequency, has until now
not improved this situation \cite{Bretin04}, although it has opened up the
possibility of forming a giant vortex in the center of the cloud
\cite{Lundh02, Kasamatsu02, Kavoulakis03}. 
Applying a one-dimensional optical lattice produces pancake condensates each
containing typically 1000 particles, such that quantum fluctuations are strongly
enhanced. Experimental signal is still conserved because of the combined
signal of all the pancake.
Moreover, because of the small number of particles in each pancake shaped
Bose-Einstein
condensate, finite-size
effects become very pronounced in this setup. In particular, 
the critical filling factor for the melting of the lattice $\nu_c$ changes
compared to the  
homogeneous situation. As a further consequence, melting is not expected to
occur homogeneously
but starts at the outside and then gradually moves inwards as the rotation speed
increases \cite{Fischer04}. Therefore,  phase coexistence is expected, where a vortex crystal is
surrounded by a vortex liquid. 
 
The optical lattice gives also the exciting possibility  to
study the dimensional crossover between two-dimensional melting and
three-dimensional melting, by varying the
coupling between the pancakes.
This system also exhibits an intruiging similarity to the layered structure of
the 
high-$T_c$ superconductors \cite{Rozhkov96, Col05}.
Recently, the density profiles for quantum Hall liquids in this geometry have
been calculated \cite{Cooper05}. 
Also the static properties of the lattice in a double and multilayer geometry
have been investigated \cite{Zhai04, Zhang05}. Finally, classical melting
between shells of vortices in a single
two-dimensional Bose-Einstein condensate has also been studied recently
\cite{Pogosov06}.

In this article we study this interesting physics by expanding
on our previous work \cite{Snoek06}.
The paper is organized as follows. In Sec. \ref{meltII} we present the theoretical
foundations of our work and derive the effective
lagrangian for the vortex fluctuations. In Sec. \ref{meltIII} we discuss the excitation
spectrum
and  we pay in particular attention to the Bloch bands of the Tkachenko modes. 
In Sec. \ref{meltIV} we present result on the quantum melting, and in Sec. \ref{meltV} we take into
account the effects of a nonzero temperature. We calculate the crossover
temperature from
quantum melting to classical melting.
In Sec. \ref{meltVI} we present results on the addition of a quartic potential. We end up
with conclusions in Sec. \ref{meltVII}.

\section{Effective Lagrangian} \label{meltII}
In this section we derive the effective lagrangian for the vortex fluctuations.
Using the ansatz that the condensates 
wavefunctions are part of the lowest Landau level, we find the classical
groundstates and expand the action up to second order in the fluctuations.

\subsection{Tight-binding approximation}
We closely follow the approach in Ref. \cite{Martikainen04} for a single vortex,
extending it to the case of a vortex lattice.
Our starting point is a cigar-shaped Bose-Einstein condensate trapped by the
potential
\begin{equation}
V({\bf r}) = \frac{m}{2}( \omega_\perp^2 r^2 + \omega_z^2 z^2),
\end{equation}
where  $m$ is the atomic mass and $\omega_r$ and $\omega_z$ are the radial and
axial trapping frequencies,
respectively. Since we assume a cigar-shaped trap, we have that
$\omega_z \ll \omega_\perp$ and we neglect the axial trapping in the rest of
this
work. Experimentally this can also be realized by using end-cap lasers,
which
results in a box-like trapping 
potential in axial direction \cite{Strecker02}. In addition, the Bose-Einstein
condensate
experiences a one-dimensional optical lattice 
\begin{equation}
V_z({\bf r}) = V_z \sin^2\left( \frac{2 \pi  z}{\lambda}\right),
\end{equation}
where $V_z$ is the lattice depth and $\lambda$ is the wavelength of the laser
light. 
The lattice potential splits the condensate into $N_s$
two-dimensional condensates with a pancake shape. Each of the pancake
Bose-Einstein condensates contains on average $N$ atoms, so the total number of
atoms is $N_s N$.
%We take the lattice to be sufficiently deep
%such that its depth is larger than the chemical potential
%of the two-dimensional condensate. 
We consider a deep optical lattice, such that its depth is larger than the
chemical potential of the two-dimensional condensate, but still we assume
that there is full coherence 
across the condensate array. This means in particular that
the lattice potential should not be so deep as to induce a superfluid-Mott
insulator
transition. Typically the required lattice depth to reach
the superfluid-Mott insulator transition in a three-dimensional lattice with 
one particle per lattice site is
of the order of $10 E_z$, where $E_z$ is the recoil energy of an atom
after absorbing a photon from the laser beam and given by
\begin{equation}
E_z=\frac{(2 \pi \hbar /\lambda)^2}{2 m}.
\end{equation}
In a one-dimensional lattice
the number of atoms in each lattice site is typically much larger 
than in a three-dimensional lattice and the transition into 
the insulating state requires a much deeper lattice \cite{Oosten03}. 
The superfluid-Mott insulator transition in a one-dimensional optical lattice
has recently been observed \cite{Stoferle04}, but also in that case
the number of particles per lattice site is of the order of one.

The action describing the system in the rotating frame is given by $S= \int dt
\int d^3 {\bf x} \; \mathcal{L}({\bf x}, t)$, where the Lagrange density is
given
by
\begin{equation}
\mathcal{L} =  \Psi^* 
\left(i \hbar \partial_t 
+ \frac{\hbar^2 \nabla^2}{2m} 
- V_{\rm ex}({\bf x}) 
+ \Omega L_z   - \frac{g}{2} |\Psi|^2 \right)\Psi.
\end{equation}
Here $\Psi ({\bf x}, t)$ is the Bose-Einstein condensate wavefunction, and
$\Omega$ is the rotation frequency. Moreover,
\begin{equation}
L_z= i \hbar (y \partial_x - x
\partial_y)
\end{equation}
is the angular momentum operator,
and $g= 4 \pi \hbar^2 a/m$ is
the interaction strength, 
with $a$  the three-dimensional scattering length. 

Since we assume a deep optical lattice potential, we can perform
a tight-binding
approximation and
write 
\begin{equation}
\Psi({\bf x},t)= \sum_i \Phi (z-z_i) \Phi_i(x,y, t),
\end{equation}
where $i$ labels the lattice sites and $z_i = i \lambda/2$ is the position of
the $i$th site.
The wavefunction in the $z$ direction $\Phi(z)$ is
chosen
to be the lowest Wannier function
of the optical lattice.  Because of the deep optical lattice this wavefunction
is well approximated by the 
ground-state wavefunction of the harmonic approximation to the lattice potential
near the lattice minimum. 
The frequency associated with this harmonic trap is
\begin{equation}
\omega_L = \frac{ 2 \pi}{\lambda} \sqrt{ \frac{2 V_z}{m}},
\end{equation}
and the wavefunction $\Phi(z)$ is given by
\begin{equation}
\Phi(z)= (\pi \ell_L^2)^{1/4} \exp \left( - \frac{z^2}{2
\ell_L^2}\right), 
\label{wavefunc_z}
\end{equation}
where $\ell_L = \sqrt{\hbar/m \omega_L}$.

Writing the Lagrange density as 
$\mathcal{L} = \mathcal{T} - \mathcal{E}$,
the time-derivative part of the Langrage density becomes
\begin{equation}
\mathcal{T} = 
\sum_i \mathcal{T}_i =  
%\sum_i \Psi_i^*  (i \hbar \partial_t) \Psi_i = 
\sum_i \frac{i \hbar}{2} \left(\Psi_i^*  \partial_t \Psi_i - \Psi_i 
\partial_t \Psi_i^* \right).
\end{equation}
Terms coupling wavefunctions on  neighboring sites do not appear in this
kinetic part of the lagrangian, because of the orthogonality of the Wannier
functions on different sites.
The energy part of the Lagrange density can be written as
\begin{equation}
\mathcal{E} = \sum_i \mathcal{E}_i + \sum_{\langle i j \rangle}
\mathcal{J}_{ij},
\end{equation}
with
\begin{eqnarray}
\mathcal{E}_i &=& \Psi_i^* \left( -\frac{\hbar^2 \nabla^2}{2m} 
+ \frac{ m \omega_\perp^2}{2}  r^2
- \Omega L_z - \frac{g'}{2} |\Psi_i|^2 \right)\Psi_i
\nonumber, \\
\mathcal{J}_{ij} &=&  -t \left( \Psi_i^* \Psi_j + \Psi_j^* \Psi_i \right), 
\end{eqnarray} 
and $\langle i j\rangle$ denotes that the sum is taken over 
nearest-neighboring sites.
We have defined the effective two-dimensional coupling strength 
\begin{equation}
g'=g \int d z |\Phi(z)|^4 = \frac{4 \pi \hbar^2 a}{\sqrt{2 \pi}\ell_L m}
\end{equation}
and the hopping amplitude
\begin{equation}
t = - \int d_z \Psi(z) \left[ - \frac{ \hbar^2 \partial_z^2}{2 m} + V_z(z)
\right]\Psi(z + \lambda/2).
\end{equation}
Using the Gaussian wavefunction from Eq. (\ref{wavefunc_z}) to calculate this
hopping amplitude
underestimates the hopping amplitude, since this approximation is only good
in the vicinity of the center 
of the wells of the optical lattice, and not in the classically forbidden
regions where the overlap between the neighboring wavefunctions is maximal.
Therefore, we use the result
\begin{equation}
t=4 V_z^{3/4} E_z^{1/4} \exp[-2 \sqrt{\frac{V_z}{E_z}}]/\sqrt{\pi}
\end{equation}
which is exact for a deep optical lattice \cite{abromowitz}.
The parameters $g'$ and $t$ are  both fully determined by the microscopic
details of the atoms and the optical lattice.

\subsection{Lowest Landau level approximation}
Melting is only expected for a Bose-Einstein condensate that is weakly 
interacting. The transverse wavefunction can then be taken to be part of the
lowest 
Landau level \cite{Ho01, Watanabe04}. This implies that we describe the Bose-Eintein condensate as a compressible fluid.
Thus we consider wavefunctions
of 
the form 
\begin{equation}
\Phi_i(x,y,t) \propto  \prod_n (w-w_{n i}(t))\exp[-|w|^2/2],
\end{equation}
where $w = (x+ i y)/\ell$, 
$w_{n i}(t) = (x_{n i} (t) + i y_{n i}(t))/\ell$ and ${\bf x}_{ni}(t)=(x_{n i}
(t),
y_{n i}(t))$ is the position of
the 
$n^{\rm th}$ vortex at site $i$. The vortex positions are the dynamical
variables and are therefore time dependent. 
Here $\ell$ is the ``magnetic length'', which is
normally identified with
the radial harmonic length $\ell_\perp = \sqrt{\hbar/m \omega_\perp}$.
The validity of the lowest Landau level approximation can then in mean-field
theory be estimated to 
be limited by the condition \cite{Stock05}
\begin{equation}
\sqrt{8 \pi} N \frac{a}{\ell_L} < \left(1 - \frac{\Omega}{\omega_\perp}\right), 
\end{equation}
which is recently compared with exact diagonalization results for small numbers
of particles \cite{Morris06}.
To increase the validity of this study, we use $\ell$ as a variational
parameter 
instead of fixed it to the radial harmonic length, such that our results are
also valid for stronger interactions and slower rotation \cite{Mueller04}. 
The associated frequency is $\omega=\hbar/m \ell^2$. In practice, it turns out
that $\omega$
is always well approximated by
the rotation frequency $\Omega$.

Making use of the fact that the wavefunctions $\phi_n (w)= {w^n}e^{-|w|^2/2}/\sqrt{\pi
\ell^2 n!}$ form a complete and 
orthonormal basis for the lowest Landau level wavefunctions, it is easy to
derive that within these approximations we have that
\begin{equation}
\int \!\! d^2 {\bf r} \; \Phi_i^* \left(-\frac{\hbar^2 \nabla^2}{2 m}\right)
\Phi_i \!\!=\!\!  
\frac{\hbar \omega}{2 \ell^2} \! \int \!\! d^2 {\bf r}
r^2  n_i({\bf r}), 
%\\
%\int d^2 \!\! {\bf r} \; \Phi_i^*\left( \frac{m}{2} \omega_\perp^2 r^2 \right)
%\Phi_i \!\!&=&\!\!  
%N \frac{\hbar \omega_\perp}{2 \omega^2} \! \int \!\! \frac{d^2 {\bf
%r}}{\ell^2} 
%\left(\frac{r}{\ell}\right)^2 \!\! n_i({\bf r}) \\
%\int \!\! d^2 {\bf r} \; |\Phi_i(x,y,t)|^4 \!\! &=& \!\!  \frac{N^2 g'}{\ell^2}
%\! \int \!\! \frac{d^2 {\bf r}}{\ell^2} n_i^2 ({\bf r}),
\end{equation}
where the density $n_i({\bf r})$ is a function of the vortex positions ${\bf
x}_{ni}(t)$.

\begin{figure*}
\begin{center}
\includegraphics[scale=1.0]{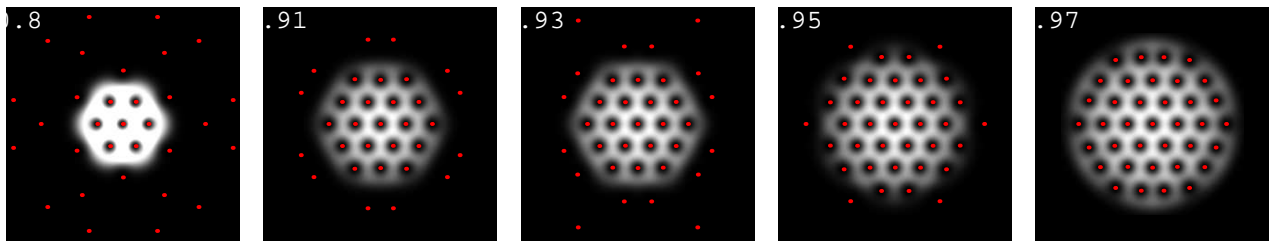}
\end{center}
\caption{(Color online) Classical vortex lattice and density profile 
for rotation frequencies $\Omega/\omega_\perp=.8$, $.91$, $.93$ $.95$ and $.97$.
Here
$U=10$, which corresponds to $N=250$.
White means high density, black low density. The vortex positions are indicated
by a dot, such that
also the vortices outside the condensate are visible.}
\label{latticel}
\end{figure*}

From now on distances are rescaled by $\ell$,
frequencies are scaled by the radial trapping frequency $\omega_\perp$, 
and we define a dimensionless interaction strength by means of
\begin{equation} 
U= N \frac{m g'}{4 \pi \hbar^2} =  N \frac{a}{\sqrt{ 2
\pi}\ell_L}.
\end{equation}
The on-site part of the energy functional can then be written as:
\begin{equation}
\frac{\mathcal{E}_i}{\hbar \omega_\perp N}
= \frac{1}{2}\left(\omega + \frac{1}{\omega} - 2 \Omega\right)
 r^2 n_i({\bf r})
+ 2 \pi \omega U  n_i^2( {\bf r}).
\label{onsiteenergy}
\end{equation}
The lowest Landau level wavefunctions and, therefore, also the atomic density,
are fully determined by the location of the vortices. To consider the quantum
mechanics of the vortex lattice we, therefore, replace
the functional integral over the condensate wavefunctions by a path integral
over the vortex positions. This involves a non-trivial Jacobian, which does not
change the results presented here, because we always consider the case that $N
\gg 1$. 
In the calculations we take the scattering length of $^{87}$Rb,
$\lambda = 700$ nm and $V_z/E_z=16$, which gives $U=25N$. The qualitative
features, however, do not depend on 
the value of $U$.

\subsection{Classical Abrikosov lattice}
\begin{figure}
\begin{center}
\psfrag{Omega}{$\Omega$}
\psfrag{N_v}{$N_v$}
\includegraphics[scale=.66]{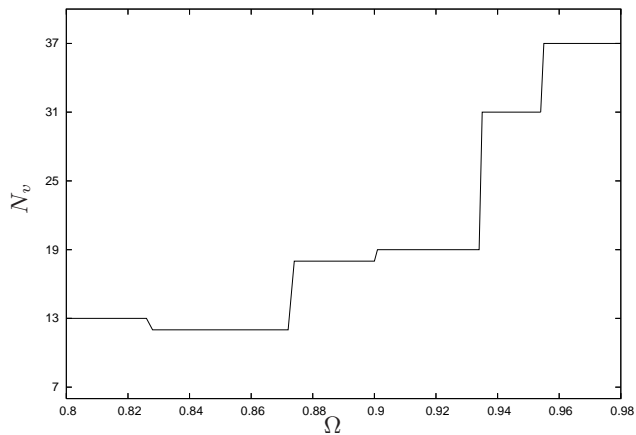}
\end{center}
\caption{The number of vortices $N_v$ that is within the condensate as a
function of the rotation frequency $\Omega$. Depending on the rotation
frequency, the groundstate is either a vortex lattice with a vortex in the
center surrounded by rings of a multiple of six vortices such that the number of
vortices can be written as $1 + 6 n_r$, or a configuration where there is no
vortex in the center and the vortices order themselves in rings of multiples of
three vortices, such that the total number of vortices in the condensate is a
multiple of three. For $\Omega/\omega_\perp>.9$, the first situation is the
groundstate (except for a small region, where the Bose-Einstein condensate becomes elliptically deformed) and we only calculate the fluctuations for this case. For this plot
the condensate radius was defined
as the radius where the angular averaged density drops below $0.003$.}
\label{fig_av}
\end{figure}

To determine the quadratic fluctuations around the Abrikosov
lattice, we first have to find the classical groundstate. We
calculated this
groundstate for up to 37 vortices. The number of vortices in the condensate 
increases with the rotation frequency.
For small numbers of vortices, the
groundstate is
distorted from the hexagonal lattice \cite{Aftalion05, Aftalion06}. In
general, there are also
vortices far outside the condensate. 
When there are more than 18 vortices in the condensate, there is generally one
vortex in the center, while
the other vortices order themselves in rings of multiples of six vortices.
However, when a new ring of vortices enters the condensate, there is an
instability towards an elliptic shape deformation. This is related to the
elliptic shape deformation that occurs before a single vortex enters the
condensate, and that has been investigated before theoretically \cite{Dalfovo01,
Recati01, Sinha01, Kramer02} and has also been observed \cite{Madison01}.
This shape instability plays an important role in the classical melting of the
vortex lattice, as we show lateron.
Pictures
of the classical groundstate are given in Fig.\ \ref{latticel} for fixed
interaction $U$ and
different rotation frequencies. 
The
number of vortices within the condensate as a 
function of rotation frequency is plotted in Fig. \ref{fig_av}, where we limited
our study to 
vortex lattices that exhibit hexagonal symmetry around the origin. In the
fluctuation calculation we only consider the regime where configurations
consisting of one vortex in the center surrounded by rings of six vortices are
stable.  
In that case, the coarse-grained atomic density is
well approximated by a Thomas-Fermi profile \cite{Cooper04}.

\subsection{Fluctuations}
Next we study the quadratic fluctuations by expanding the action up to second
order in the fluctuations ${\bf u}_{ni} = {\bf x}_{ni} - \langle {\bf x}_{ni}
\rangle$.
This yields an action of the form
\begin{equation}
S=\sum_i {\bf u}_i \cdot ({\bf T} i \partial_t - {\bf E}) \cdot {\bf u}_i - t
\sum_{\langle i
j \rangle} {\bf u}_i \cdot {\bf J} \cdot {\bf u}_j,
\end{equation} 
where ${\bf u}_i \equiv(\dots, {\bf u}_{ni}, \ldots)$ is the total displacement
vector of all the point vortices on site $i$. 
The matrices $\bf T$, $\bf E$ and $\bf J$
depend on $\Omega$, $U$, and the classical lattice positions
$\langle {\bf x}_{ni}\rangle$, and are formally given by 
\begin{eqnarray}
{\bf E}_{n m} &=& \left( \begin{array}{cc} E_{x_{ni} x_{mi}} & E_{x_{ni}
y_{mi}} 
\\ 
E_{y_{ni} x_{mi}} & E_{y_{ni} y_{mi}} \end{array} \right)
\nonumber \\
&=&
\int d^2 {\bf r} 
\frac{\partial}{\partial {\bf x}_{ni}}\frac{\partial}{\partial {\bf x}_{m i}}
\mathcal{E}_i,
 \nonumber \\
{\bf T}_{n m} &=& \int d^2 {\bf r} 
\frac{\partial}{\partial {\bf x}_{ni}}\frac{\partial}{\partial {\bf x}_{m i}}
\mathcal{T}_i,
\\
{\bf J}_{n m} &=&\int d^2 {\bf r} 
\frac{\partial}{\partial {\bf x}_{ni}}\frac{\partial}{\partial {\bf x}_{m j}}
\mathcal{J}_{ij}. \nonumber
% \left( \begin{array}{cc} 
%	\partial_{x_{n i}} \partial_{x_{m j}} \mathcal{J}_{ij} & 
%	\partial_{x_{n i}} \partial_{y_{m j}} \mathcal{J}_{ij} \\
%	\partial_{y_{n i}} \partial_{x_{m j}}\mathcal{J}_{ij} &
%	\partial_{y_{n i}} \partial_{Y_{m j}} \mathcal{J}_{ij}
%\end{array}
%\right). 
\end{eqnarray}
We also expand the density in the fluctuations by means of
\begin{equation}
n_i ({\bf r}) = \frac{n_0  ({\bf r}) + \sum_n {\bf u}_{n i} \cdot {\bf n}_{n}
({\bf r}) +
\sum_{n m}  {\bf u}_{n i} \cdot {\bf n}_{n m}  ({\bf r}) \cdot {\bf u}_{m i}}
{1   + \sum_n {\bf u}_{n i} \cdot {\bf n}_{n i}  + \sum_{n m}  {\bf u}_{n i}
\cdot {\bf n}_{n m} \cdot {\bf u}_{m i}  },
\end{equation}
where $n_0  ({\bf r})$ is the equilibrium particle density associated with the
classical
lattice.
We have defined the vectors
\begin{equation}
{\bf n}_n ({\bf r}) = \left( \begin{array}{c} n_{x_n} ({\bf r})\\ n_{y_n} ({\bf
r})\end{array} \right),
\end{equation}
which form the dipole densities and are associated with linear variations around
the classical solution, and the tensors
\begin{equation}
{\bf n}_{n m} ({\bf r})= \left( \begin{array}{c c} n_{x_n x_m} ({\bf r}) &
n_{x_n y_m} ({\bf r}) 
\\ n_{y_n x_m} ({\bf r}) & n_{y_n y_m} ({\bf r}) \end{array} 
\right),\\
\end{equation}
which form the quadrupole densities
and are associated with the  quadratic
variations around the classical solution.
Moreover,
\begin{equation}
{\bf n}_{n} = \int d^2  {\bf r}  \; {\bf n}_{n} ({\bf r})
\end{equation}
and 
\begin{equation}
{\bf n}_{n m} = \int d^2  {\bf r} \;  {\bf n}_{n m} ({\bf r}) ,
\end{equation}
such that $n_i ({\bf r})$ is always normalized to 1. The density $n_0 ({\bf
r})$, 
and the tensors ${\bf n}_n ({\bf r})$, ${\bf n}_{nm} ({\bf r})$ 
are independent of time and the layer index $i$, since only the fluctuations in
the vortex positions ${\bf u}_i (t)$
are taken as
the dynamic variables.
Making use of the fact that
\begin{equation}
n_i({\bf r}) \propto \prod_i |w - w_i|^2 \exp[- |w|^2],
\end{equation}
the following expressions can be
derived for the dipole density
\begin{eqnarray}
n_{x_n} ({\bf r}) &=& \frac{- 2 (x-\langle x_{n i} \rangle)} {|{\bf r} - \langle
{\bf x}_{n i} \rangle |^2} n_0
({\bf r}), \\
n_{y_n} ({\bf r}) &=& \frac{- 2 (y-\langle y_{n i} \rangle )} {|{\bf r} -\langle
{\bf x}_{n i} \rangle |^2} n_0
({\bf r}). \nonumber
\end{eqnarray}
For the quadrupole density for a single vortex we get
\begin{eqnarray}
n_{x_n x_n} ({\bf r}) &=& \frac{1} {|{\bf r} - \langle {\bf x}_{n i} \rangle
|^2} n_0 ({\bf
r}),\\ 
n_{x_n y_n} ({\bf r}) &=& 0, \nonumber\\ 
n_{y_n x_n} ({\bf r}) &=& 0, \nonumber\\
n_{y_n y_n} ({\bf r}) &=& \frac{1} {|{\bf r} - \langle {\bf x}_{n i} \rangle
|^2} n_0 ({\bf r}) \nonumber,
\end{eqnarray}
whereas for two different vortices we find
\begin{eqnarray}
n_{x_n  x_m} ({\bf r}) &=& \frac{2 (x- \langle x_{n i} \rangle)(x- \langle x_{m
i} \rangle)} 
{|{\bf r} - \langle {\bf x}_{n i} \rangle |^2 |{\bf r} - \langle {\bf x}_{m i}
\rangle |^2} n_0 ({\bf r}),\\ 
n_{x_n y_m} ({\bf r}) &=& \frac{2 (x- \langle x_{n i} \rangle)(y- \langle y_{m
i} \rangle)} 
{|{\bf r} - \langle {\bf x}_{n i} \rangle |^2 |{\bf r} - \langle {\bf x}_{m i}
\rangle |^2} n_0 ({\bf r}),\nonumber\\ 
n_{y_n x_m} ({\bf r}) &=& \frac{2 (y- \langle y_{n i} \rangle)(x- \langle x_{m
i} \rangle)}
{|{\bf r} - \langle {\bf x}_{n i} \rangle |^2 |{\bf r} - \langle {\bf x}_{m i}
\rangle |^2} n_0 ({\bf r}),\nonumber\\ 
n_{y_n y_m} ({\bf r}) &=& \frac{2 (y- \langle y_{n i} \rangle)(y-  \langle y_{m
i} \rangle)}
{|{\bf r} - \langle {\bf x}_{n i} \rangle |^2 |{\bf r} - \langle {\bf x}_{m i}
\rangle |^2} n_0 ({\bf r}). \nonumber
\end{eqnarray}
All matrices in the action for the fluctuations can be completely expressed in
terms of these functions. From Eq.
(\ref{onsiteenergy}) we read off that
\begin{widetext}
\begin{eqnarray}
{\bf E}_{n m}  &=& \int d^2 {\bf r} 
\left\{
	\frac{1}{2}(\omega + 1/\omega - 2 \Omega)r^2 
	\Bigl[ 
( {\bf n}_{n} {\bf n}_{m} - {\bf n}_{nm}) n_0({\bf r})
- \frac{1}{2} ({\bf n}_{n} {\bf n}_{m}({\bf r}) + {\bf n}_{m} {\bf n}_{n}({\bf
r})) 
+ {\bf n}_{n m}({\bf r})
	\Bigr] 
\right. 
\\
&& 
+ 2 \pi \omega U \Bigl[
(3 {\bf n}_{n} {\bf n}_{m} 
- 2 {\bf n}_{n m} n_0^2({\bf r}) - 2 \left( {\bf n}_{n} {\bf n}_{m} ( {\bf r})
n_0 ( {\bf
r})+{\bf n}_{m} {\bf n}_{n} ( {\bf r}) n_0 ( {\bf r}) \right)+ {\bf n}_{n} (
{\bf
r}) {\bf n}_{m} ( {\bf r}) + 2 {\bf n}_{n m} ( {\bf r}) n_0 ( {\bf r})
\Bigr]  \biggr\}. \nonumber
\end{eqnarray}
\end{widetext}
From a straightforward derivation we obtain also
\begin{eqnarray}
T_{x_n x_m} &=& \frac{n_{y_n} n_{x_m} - n_{x_n} n_{y_m}}{4} + \frac{n_{y_n x_m}
- n_{x_n y_m}}{2}, \\ 
T_{x_n y_m} &=& \frac{n_{x_n} n_{x_m} + n_{y_n} n_{y_m}}{4} - \frac{n_{x_n x_m}
+ n_{y_n y_m}}{2}, \nonumber\\
T_{y_n x_m} &=& - T_{x_n y_m}, \nonumber\\ 
T_{y_n y_m} &=& T_{x_n x_m},\nonumber
\end{eqnarray}
and
\begin{eqnarray}
J_{x_n x_m} &=& -\frac{n_{x_n} n_{x_m}} {4} + \frac{n_{x_n x_m} + n_{y_n
y_m}}{2}, \\ 
J_{x_n y_m} &=& -\frac{n_{x_n} n_{y_m}} {4} + \frac{n_{x_n y_m} - n_{y_n
x_m}}{2}, \nonumber\\ 
J_{y_n x_m} &=& -\frac{n_{y_n} n_{x_m}} {4} + \frac{n_{y_n x_m} - n_{x_n
y_m}}{2}, \nonumber\\ 
J_{y_n y_m} &=& -\frac{n_{y_n} n_{y_m}} {4} + \frac{n_{x_n x_m} + n_{y_n
y_m}}{2}.\nonumber 
\end{eqnarray}
The calculation is simplified considerably by making use of the hexagonal
symmetry of the
vortex lattice. Because of this symmetry the number of matrix elements to be
calculated for each matrix is less
then $12 N_v^2 + 32 N_v$ instead of $72 N_v^2$, when all matrix elements are
calculated independently.

\subsection{Diagonalization}
To diagonalize this action along the $z$ axis, we perform a Fourier
transformation to obtain
\begin{equation}
S=\sum_k {\bf u}_k^*  \cdot \left({\bf T} i \partial_t - {\bf E} - t \lbrack
1-\cos (k
\lambda/2) {\bf J} \rbrack  \right) \cdot {\bf u}_k.  
\end{equation} 
Finally, we completely diagonalize this action by a transformation 
\begin{equation}
{\bf v}_k={\bf P}_k {\bf u}_k,
\end{equation}
that is normalized such that the action becomes
\begin{equation}
S=\sum_{k, \alpha} v_{k\alpha}^*  (i \partial_t - \omega_\alpha(k) )  v_{k
\alpha},  
\end{equation}
where $\omega_\alpha(k)$ are the mode frequencies of the vortex lattice. This
means that the $v_{k\alpha}$, where $k$ labels the momentum in the $z$ direction
and $\alpha$ labels the mode, correspond to bosonic
operators 
with commutation relation $[v_{k\alpha}, {v_{k'\alpha'}}^\dagger] = \delta_{kk'}
\delta_{\alpha \alpha'}$. This allows us to calculate the expectation value for
the
fluctuations in the vortex positions, but also for 
the correlations between the various point vortices. 

\section{Tkachenko modes} \label{meltIII}
\begin{figure}
\begin{center}
\psfrag{omega(k)}{$\omega(k)/\omega_\perp$}
\psfrag{k}{$k \lambda/2$}
\includegraphics[scale=.66]{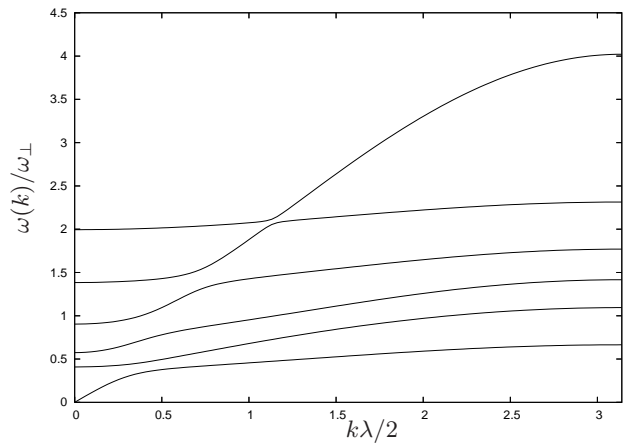}
\end{center}
\caption{Dispersion of the Tkachenko modes along the direction of the optical
lattice. For this plot the parameters were chosen as: $U=10$, $t=1/10$, and
$\Omega
= .97$. One gapless linear mode and five
gapped tight-binding-like modes can be identified. The gapless mode is the
acoustic Goldstone
mode associated with the broken $O(2)$-symmetry due to the presence of the
vortex lattice. Note that these results were obtained in the lowest Landau level approximation, which corresponds to the compressible limit.
}
\label{dispersiontkl}
\end{figure}

\begin{figure}
\begin{center}
\includegraphics[scale=1.2]{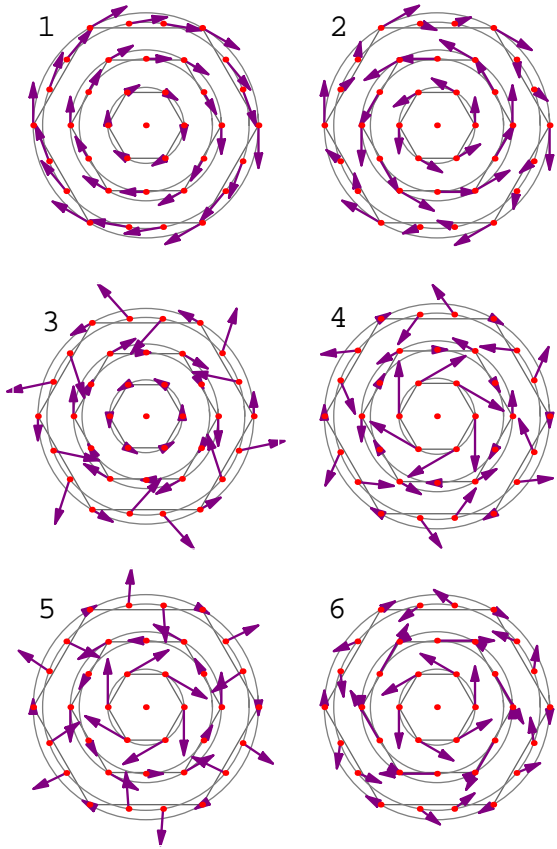}
\end{center}
\caption{(Color online) Tkachenko modes. For this plot the parameters were
chosen as: $U=10$,
$\Omega
= .97$, and $k=0$. Mode 1 is the Goldstone mode and corresponds to a pure
rotation, while the modes 2-6 are gapped and ordered along increasing gap. They
were previously called $s$-band modes in Ref. \cite{Coddington03}. As is
visible, some
of the vortices at the edge deviate from having a purely angular motion. 
}
\label{modes1}
\end{figure}

\begin{figure}
\begin{center}
\includegraphics[scale=.75]{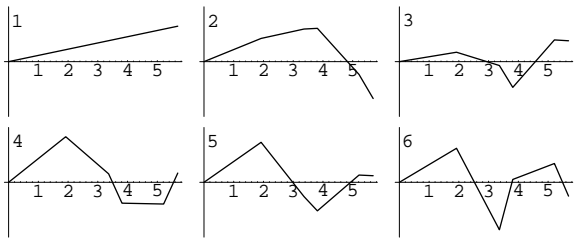}
\end{center}
\caption{Node structure of the Tkachenko modes. The modes are ordered in the
same way as in Fig. \ref{modes1}.  Points are connected by straight lines to
increase the visibility. When a wavelength is associated with the position of
the nodes, it can be seen that for increasing gap the wavelength becomes
shorter. This agrees qualitatively with the continuum theory of Ref.
\cite{Sonin05}.}
\label{modes2}
\end{figure}

The Tkachenko modes are almost transverse modes of the vortex lattice.
In a harmonic trap with cylindrical symmetry they 
become modes which are almost angular.  In the radial direction their spectrum
is discretized, because of the finite lattice size. The number of radial
Tkachenko modes equals the number of rings in which the vortices have ordered themselves. For 37 vortices 6 Tkachenko
modes can be identified. A close comparison with continuum theory for a
finite-size system, where also a discrete spectrum was found \cite{Sonin05}, is
possible but beyond the scope of this work.
Moreover, the Tkachenko modes also have a dispersion in the $z$ direction.
Without the optical lattice some aspects of these modes were recently
investigated \cite{Chevy05}.
For typical parameters this dispersion is plotted in Fig. \ref{dispersiontkl},
while the modes are displayed in Fig. \ref{modes1}. As can be seen in the latter
figure,
some of the vortices at the edge deviate from having a purely angular motion.
As is clearly visible, there is one gapless mode, which is linear at long
wavelengths, while the other modes 
are roughly just tight-binding-like. Moreover, various avoided crossings
between these modes are clearly visible. The gapless mode is the Goldstone mode
associated with the spontaneously broken rotational $O(2)$-symmetry due to the
presence of the vortex lattice. 
When the tunneling rate is very small, the gapped modes have exactly a
tight-binding dispersion and the gapless mode gets a dispersion proportional to
$\sin(k \lambda /4)$. This can be understood by observing that in this case the
modes are decoupled and the hamiltonian for the gapless mode reduces to the
Josephson hamiltonian
\begin{equation}
\mathcal{H}= - E_C \sum_i \frac{\partial^2}{\partial \phi_i^2} +
E_J \sum_{\langle i j \rangle} \cos(\phi_i - \phi_j)^2.
\label{goldstone}
\end{equation}
Writing $p_\phi = - i \partial_\phi$, the action in momentum space thus reads 
after a
quadratic expansion of the Josephson energy
\begin{equation}
\mathcal{S} =  \sum_k \int d \omega  \left( \! \begin{array}{c} \phi \\ p_\phi
\end{array} \! \right) 
\left( \! \begin{array}{cc} 2 \lbrack 1 - \cos (k \lambda/2)\rbrack  E_J & - i
\omega \\i \omega & E_C  \end{array} \! \right)
\left( \! \begin{array}{c} \phi \\ p_\phi \end{array} \! \right),
\end{equation}
from which we deduce the dispersion
\begin{equation}
\omega^2(k) = 2 \lbrack 1 - \cos (k \lambda/2)\rbrack  E_J E_C = 4 \sin^2 ( k
\lambda/4) E_J E_C.
\end{equation}

It is interesting to note that for a small rotation frequency, which implies a
small vortex lattice, the Tkachenko
modes are not the lowest-lying modes. For $U=10$, a Tkachenko
mode becomes the lowest-lying gapped mode when $\Omega> 0.978$, but there are
many
modes in between the second and the third Tkachenko mode.
This confirms the expectation that increasing the vortex lattice will bring down
the Tkachenko spectrum more and more.

\section{Quantum melting} \label{meltIV}
In this section we study vortex-lattice melting due to quantum
fluctuations. 
We apply the Lindemann criterion to estimate the position of the melting
transition.
We study the influence of tunneling between the pancake condensates and compare
with a local-density approximation. By looking at correlations between the
vortices, we can distinguish between various phases, which are partially
melted.

\subsection{Single-layer geometry}
Quantum fluctuations of the vortices ultimately result in
melting of the vortex lattice.
To decide whether or not the lattice is melted, we use the Lindemann criterion,
which in this inhomogeneous situation has to be applied
locally.
The Lindemann criterion means that the lattice is melted, when
\begin{equation} 
\frac{\langle {\bf u}_{ni}^2 \rangle}{\Delta_{ni}^2} > c_L^2,
\end{equation} 
where  $\Delta_{ni}$ is  average distance to the
neighboring vortices.  
The critical value $c_L$, is known as the Lindemann parameter. This
parameter is {\it a priori} unknown, but values found from comparison with Monte
Carlo
simulations are typically in the range $c_L=0.1-0.3$.
The value  
$c_L =0.1$ was recently found from elaborate
calculations for a triangular vortex lattice in high-$T_c$
superconductors, which also compared well to experiments in that case
\cite{Dietel05}. We, therefore, use this
value in our calculations. Note that  the results  we obtain depend
quantitatively
on the value of the Lindemann parameter. Changing this value shifts the curves,
but the qualitative features
remain the same.

\begin{figure}
\begin{center}
\psfrag{N_v}{$N_v$}
\psfrag{Omega}{$\Omega/\omega_\perp$}
\psfrag{R_cr/R}{ $R_{\rm cr} / R$}
\includegraphics[scale=.66]{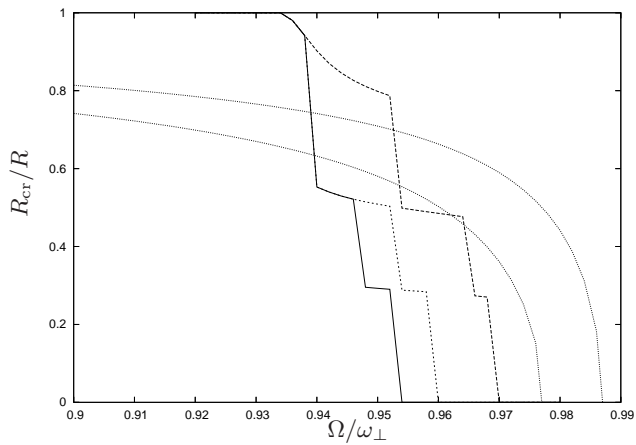}
\end{center}
\caption{Crystal radius $R_{\rm cr}$ normalized to the condensate radius $R$ as
a function
of the rotation frequency for $N= 250$ and $U=10$. The solid line is for $t=0$
and the
dashed lines are for
$t=1/1000$  and $1/100$ (from left to right).
The dotted lines  are the result of the local-density
approximation, where for the upper curve we took the criterion $N/N_v=6$ and for
the lower curve $N/N_v=8$. In all the figures in this paper we have connected
the points by straight lines. Due to the finite numerical resolution the jump
between the plateau's appears therefore not vertical.}
\label{meltingl}
\end{figure}

Due to the inhomogeneity of our system, we have to apply this criterion
locally. 
Because the coarse-grained particle density decreases with the distance to the
origin, 
vortices on the outside are already melted, while the inner part of the crystal
remains solid \cite{Fischer04}. 
Therefore, a crystal phase in the inside  coexists with a liquid
phase on the outside. In Fig. \ref{meltingl} we compute the radius of the crystal
phase $R_{\rm cr}$ normalized 
to the condensate radius $R$, as a function of the rotation frequency for fixed
a number of particles and a fixed interaction strength $U$, but for various
hopping
strengths $t$. 
Also here we define the condensate radius $R$ as the radius for which the angularly
averaged density drops below $0.003$. The crystal radius $R_{\rm cr}$ is defined
as the radius of the innermost ring of vortices that is melted according to the
Lindemann criterion. When according to this definition  $R_{\rm cr}>R$, we set
the crystal radius equal to the condensate radius, i.e., $R_{\rm cr}=R$.  
The ratio $R_{\rm cr}/R$ shows discrete steps because of the ring-like structure
in which the vortices order themselves. 

\subsection{Local-density approximation}
We compare this with a simple local density calculation, where the criterion
$N/N_v=n(r)/n_v( r)=6$ \cite{Cooper01} or 
$N/N_v=n(r)/n_v( r)=8$  \cite{Sinova02}
is applied
locally, by making use of a Thomas-Fermi density profile to describe the
coarse-grained
atomic density. 
%In previous work \cite{Snoek06}, we compared with the value
%$N/N_v=n(r)/n_v( r)=8$ from Ref. \cite{Sinova02}, but since this value is also
%based on a Lindemann criterion, we feel it is better to compare with the exact
%diagonalization results of Ref. \cite{Cooper01}.
Substituting the Thomas-Fermi profile
\begin{equation}
n_{\rm TF}(r) = \frac{2}{\pi R_{\rm TF}^2} \left(1 - \frac{r^2}{R_{\rm TF}^2}
\right),
\end{equation}
where $R_{\rm TF}$ is the Thomas-Fermi radius, in the on-site energy in Eq.
(\ref{onsiteenergy}), we 
obtain
\[
 \left(\omega + \frac{1}{\omega} - 2 \Omega\right)\frac{R_{\rm TF}^2}{6} +
\frac{\beta 8 \omega U}{3 R_{\rm TF}^2}.  
\]
We have introduced here also the Abrikosov parameter $\beta={\int n^2({\bf
r})}/{(\int n({\bf
r}))^2}  \simeq 1.1596$  for the hexagonal lattice.
Minimizing for $R_{\rm TF}$ gives 
\begin{equation}
R_{\rm TF}^4 = \frac{ 16 \beta U \omega^2}{1+ \omega^2 - 2 \omega \Omega},
\end{equation}
which on substitution gives for the on-site energy
\[
\frac{4}{3} \sqrt{\beta U} \sqrt{1+ \omega^2 - 2 \omega \Omega}.
\]
This is minimized for
\begin{equation}
\omega= \Omega,
\end{equation}
which sets the variational parameter $\ell$.
The Thomas-Fermi radius is then given by
\begin{equation}
R_{\rm TF}^4=16  \frac{\Omega^2 \beta U}{ 1 -  \Omega^2}.
\end{equation}
In lowest order the vortex density is given by
\begin{equation} 
n_v ({\bf  r}) = \frac{1}{\pi \ell^2}.
\end{equation}
Solving now
\begin{equation}
\nu ({\bf r}) = \frac{n({\bf r})}{n_v ({\bf r})}=\nu_c,
\end{equation}
we get for the crystal 
radius normalized to the Thomas Fermi radius the result
\begin{equation}
\frac{R_{\rm cr}}{R_{\rm TF}} = \sqrt{1 - \frac{\nu_c R^2}{2 N}} = \sqrt{1-
\frac{2 \nu_c}{N} \sqrt{ 
\frac{\Omega^2 \beta U}{(1- \Omega^2)}}}.
\end{equation}
For comparison this line is plotted in Fig.
\ref{meltingl}. 
As can be seen there are important finite-size corrections to the local-density
approximation. The local-density approximation fails to take into account the
discrete nature of the vortex positions. Moreover, 
it predicts the total melting of the crystal at considerable higher rotation
frequencies than the exact answer.
%As we will show in more detail in Section V, this is in
%agreement with the fact that for the infinite crystal there is no divergence.
% This is in sharp contrast with the situation for nonzero temperature, where
%the
%long wavelength fluctuations give a divergent contribution and the fluctuations
%are smaller for a finite system.

In principle there are corrections to the vortex density due to the Thomas-Fermi
profile of the particle density. A better approximation near the center of the
trap is \cite{Ho01}
\begin{equation}
n_v ( r) = \frac{1}{\pi \ell^2} - \frac{1}{\pi R_{\rm TF}^2}
\frac{1}{(1-r^2/R_{\rm TF}^2)^2}.
\end{equation}
However, this vortex density becomes zero well within the Thomas-Fermi radius.
As a result the ratio
$n({\bf r})/n_v({\bf r})$ diverges and is always bigger than 8. Higher-order
contributes should be taken into account to solve this problem. 
Instead we compare with the constant vortex density
\begin{equation}
n_v ( r) = \frac{1}{\pi \ell^2} - \frac{1}{\pi R_{\rm TF}^2},
\end{equation}
which corresponds to a Gaussian distribution of the atomic density with radius
$R_{\rm TF}$.
Straightforward derivation gives then
\begin{eqnarray}
\frac{R_{\rm cr}}{R_{\rm TF}} &=& \sqrt{1 - \frac{\nu_c (R^2-1)}{2 N}} \\
&=& \sqrt{1-
\frac{2 \nu_c}{N}\left( \sqrt{ 
\frac{\Omega^2 \beta U}{(1- \Omega^2)}}-1 \right)}.
\end{eqnarray}
This shifts the curves for the local-density theory in Fig. \ref{meltingl} a
little bit upwards, and makes the comparison even less favourable.

\subsection{Multi-layer geometry}
\begin{figure}
\begin{center}
\psfrag{t}{$t$}
\psfrag{R_cr/R}{ $R_{\rm cr} / R$}
\includegraphics[scale=.66]{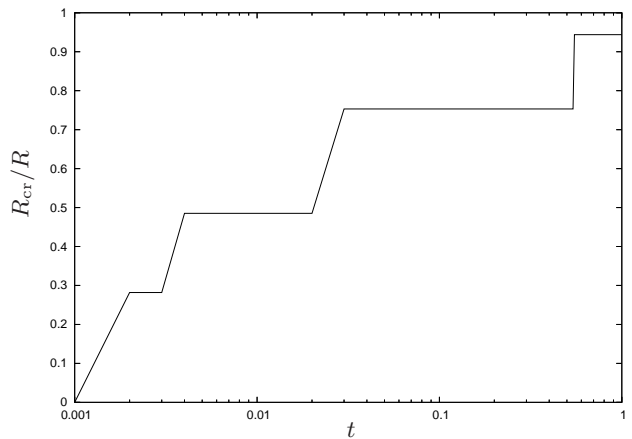}
\end{center}
\caption{Crystal radius $R_{\rm cr}$ normalized to the condensate radius $R$ as
a function 
of the hopping amplitude $t$ for $N= 250$, $U=10$ and $\Omega=.96$. }
\label{freezing}
\end{figure}

When the tunneling between pancakes is turned on, the fluctuations are also 
coupled in the  axial direction. This decreases the fluctuations in the vortex
displacements because the stiffness of the vortices increases. Hence melting
occurs for higher rotation frequencies,
as is visible in Fig. \ref{meltingl}.
In Fig. \ref{freezing} we show the crystal radius for fixed rotation frequency
and increasing hopping amplitude.
\begin{figure}
\begin{center}
\psfrag{exp}{$e^{ - \langle ({\bf u}_{n i} - {\bf u}_{n j})^2 \rangle}$}
\psfrag{z}{$2 z/\lambda$}
\psfrag{t}{$t$}
\includegraphics[scale=.66]{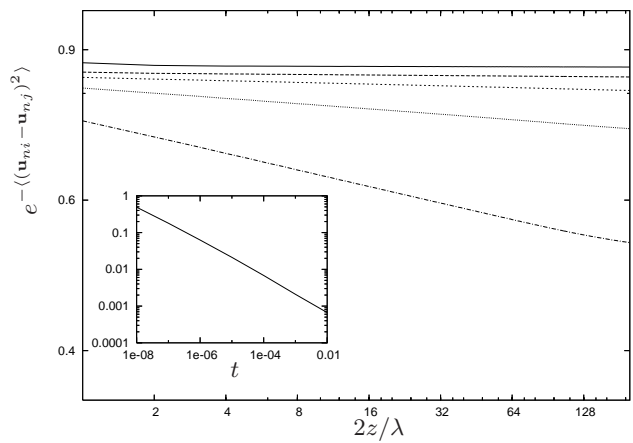}
\end{center}
\caption{Axial correlation function $e^{ - \langle ({\bf q} \cdot ({\bf u}_{n i}
- {\bf u}_{n
j}))^2 \rangle/2 }$ 
as 
a function of the distance along the axial direction
for ${\bf q}= (1/\ell, 1/\ell)$, $N= 250$, $U=10$, and
$\Omega=0.97$. The various curves have, from top to bottom, a hopping parameter
$t=10^{-2}$, $t=10^{-3}$,
$t=10^{-4}$, $t=10^{-5}$, and $t=10^{-6}$, respectively. The correlation
function decays as a power law and in the
inset the power is plotted as a function of 
the hopping amplitude $t$. The power thus scales with $1/\sqrt{t}$.}
\label{corr}
\end{figure}
To determine the presence of crystalline order in the axial direction, we
calculate the correlation function
\[
\langle e^{ i {\bf q} \cdot ({\bf u}_{n i} - {\bf u}_{n j})} \rangle,
\]
which is
related to the static structure
factor. 
For our gaussian theory, this reduces to
\[
e^{ -   \langle \lbrack {\bf q } \cdot ({\bf u}_{n i} - {\bf u}_{n j}) \rbrack^2
\rangle/2}. 
\]
For the central vortex the correlation function  $\langle {\bf u}_{0 i} {\bf
u}_{0 j} \rangle$ decays
exponentially, which signals long-range order. 
For the other vortices  $\langle \lbrack {\bf q } \cdot ({\bf u}_{n i} - {\bf u}_{n j})
\rbrack^2 \rangle$ grows as a logarithm, such
that $e^{ -  \langle \lbrack {\bf q} \cdot({\bf u}_{n i} - {\bf u}_{n j})
\rbrack^2 \rangle}$  decays
algebraically.
 This is in agreement with the
expectation for a one-dimensional system at zero temperature.
In Fig. \ref{corr} we show the axial correlation function for one of the
vortices outside the center.

Since the lattice spacing $\lambda/2$ is the only length scale in the 
$z$ direction, it also determines the axial correlation length.  
The power of the algebraic decay scales with $1/\sqrt{t}$. This can be
understood when we assume that
the gapless Tkachenko mode gives the dominant contribution to the axial
correlation.
Using the effective hamiltonian for this mode from Eq. (\ref{goldstone}), we
obtain the following expression: 
\begin{eqnarray}
&& \langle ({\bf u}_{in}-{\bf u}_{jn})^2 \rangle 
\simeq  \frac{ r_n^2}{\sqrt{E_J E_C}} \times
\\ && \frac{2 }{\pi}
 \int_0^{\pi}  d k  \sin^2 (k (i-j)/2)) 
\left(\frac{E_C}{ \sin ( k /2)} + E_J \sin (k /2) \right), \nonumber 
\end{eqnarray}
where we rescaled $k$ ,  in units of the lattice spacing $\lambda/2$, $r_n$ is
the distance of the vortex to the origin and we used that $ 1- \cos( k (i-j))= 2
\sin^2 (k (i-j)/2)$. Only the first term in the brackets is divergent for $k
\rightarrow 0$, so we neglect the second term. We make the approximation
$\sin(k/2) \simeq k/2$.  When $i-j$ is large we can approximate the rapidly
oscillating function  $\sin^2(k(i-j)/2)$ by its average $1/2$, except
for a region near $k=0$. We obtain then
\begin{eqnarray}
&&  \frac{ 4 r_n^2}{\pi} \sqrt{\frac{E_C}{E_J}} \int_0^\pi d k
\frac{\sin^2(k(i-j)/2) }{k}
\label{corrfun}
\\
&& =  \frac{ 4 r_n^2}{\pi} \sqrt{\frac{E_C}{E_K}} \left(
\int_0^{\tfrac{4 \pi}{i-j}} d k \frac{\sin^2(k(i-j)/2)} {k} +  \frac{1}{2}
\int_{\tfrac{4\pi}{i-j}}^\pi  \frac{1}{k}  \right) \nonumber \\
&&= \frac{ 4 r_n^2}{\pi} \sqrt{\frac{E_C}{E_J}} \left( 
\int_0^{4\pi} dk \frac{\sin^2 }{k} + \frac{1}{2} \log 4(i-j) \right) \nonumber\\
&&=  \frac{ 2 r_n^2}{\pi}\sqrt{\frac{E_C}{E_J}}\left( \log 4(i-j) + C_1 \right),
\nonumber
\end{eqnarray}
where the constant $C_1 = \gamma + \log( 8 \pi) + {\rm Ci}(8 \pi) \simeq
3.80296$. 
Using now the fact that $E_J \propto t$, we conclude that indeed 
 the power of the algebraic decay of $e^{ - 
\langle({\bf q} \cdot ({\bf u}_{n i} - {\bf u}_{n j}))^2 \rangle/2}$ scales with
$1/\sqrt{t}$.

For nonzero temperatures we obtain a linear rise
of the correlation function $\langle ( {\bf u}_{n i} - {\bf u}_{n j})^2\rangle$,
which
can be understood from the same argument, since then we have to add the
Bose-Einstein distribution, which 
for very low temperatures can be approximated by
\begin{equation}
\frac{1}{e^{\omega(k)/k_B T}-1} \simeq \frac{ k_B T}{\omega(k)} = \frac{ k_B T
}{2 \sqrt{E_J E_C} \sin (k \lambda /4)}.
\end{equation}
When we again rescale the momentum by the lattice constant $\lambda/2$ and
approximate $\sin k/2 \simeq k/2$, we observe that 
the $1/k$-factor in the intergrand of Eq. (\ref{corrfun}) is replaced by
$1/k^2$. This gives that up to a constant 
$\langle ({\bf u}_{in}-{\bf u}_{jn})^2 \rangle \propto  i-j$ and  $e^{ - 
\langle({\bf q} \cdot ({\bf u}_{n i} - {\bf u}_{n j}))^2 \rangle/2}$ decays
exponentially.

\subsection{Correlations}
Melting for small arrays of electrons in quantum dots that are composed of two
or three shells \cite{Bedanov94, Schweigert95,  Belousov98, Filinov01}, and 
of vortex lattice shells \cite{Lozovik98, Pogosov06} is predicted to occur in
two stages.  First the oriental order between different shells is destroyed,
while after that the radial order is destroyed. We also find this behavior for
the vortex lattice. For that we compute the correlation function between various
vortices and decompose it in 
radial and angular fluctuations. A shell structure is found, where vortices
with almost (but not necessarily exactly) the same distance group together.
Between these shells angular fluctuations dominate, while within the shell
angular fluctuations are suppressed and radial fluctuations dominate.
This can
be understood if we assume that the Tkachenko modes 
dominate the fluctuations, because they leave the rings intact, but change the
relative angle between the rings. 
\begin{figure}
\begin{center}
\includegraphics[scale=.9]{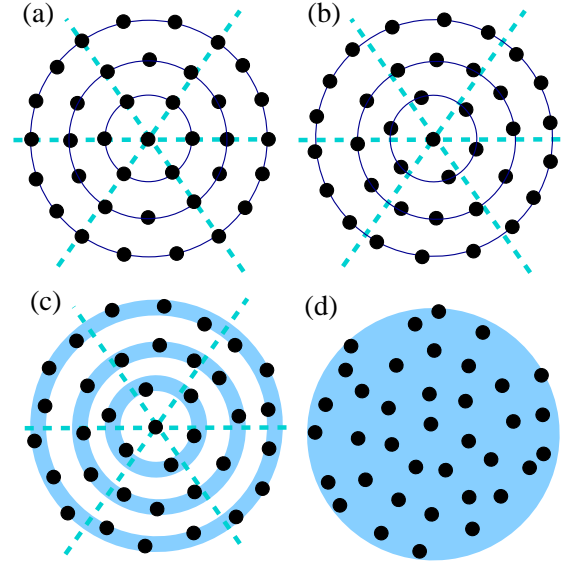}
\end{center}
\caption{(Color online) Phases that can be distinguished by looking to radial
and angular correlations between vortices on the same and on different shells:
a) a solid, b) a solid of solid shells,  c) a hexatic liquid, d) a liquid.}
\label{phases}
\end{figure}
To quantify this, we introduce the following order parameter
to measure fluctuations in the relative angle of neighboring shells
\begin{equation}
\Gamma_{nm}^{\rm SS} = \left(\frac{{\bf u}_{ni}^{\rm ang}}{r_n} - \frac{{\bf
u}_{mi}^{\rm ang}}{r_m} \right)^2,
\end{equation}
for neighboring vortices $n$ and $m$ on different shells, where $r_n$ is the
distance of vortex $n$ to the origin and 
only the angular part of the fluctuations in the displacement field is
considered. 
Let $N_v^{\rm max}$ be the maximum number of vortices on either of the two
shells. The shells are decoupled with respect to each other if 
\begin{equation}
\Gamma_{nm}^{\rm SS} > \frac{2 \pi}{N_v^{\rm max}} c_L^2, 
\end{equation}
with $c_L=0.1$ the same Lindemann parameter.
Furthermore we introduce the correlation function
\begin{equation}
\Gamma_{nm}=\frac{ \langle ({\bf u}_{ni} - {\bf u}_{ m i})^2
\rangle}{2 \Delta_{mn}^2}, 
\end{equation}
where $\Delta_{mn}$ is the distance between the neighboring vortices $m$ and
$n$ and we split this up in a radial and angular component.
Moreover we distinguish between the case when $n$ and $m$ are on the same shell,
which we denote by $\Gamma_{r}^{\rm rad}$ and $\Gamma_{r}^{\rm ang}$, or on
different shells, denoted by
$\Gamma_{r r'}^{\rm rad}$ and $\Gamma_{r r'}^{\rm ang}$.
We use the definition that order is destroyed if the
order parameters exceeds the Lindemann parameter, i.e., $\Gamma >
c_L^2=0.01$.
When the fluctuations are small, the vortex lattice is in the crystal phase.
This phase is defined as
$\Gamma_{nm} = \Gamma_{nm}^{\rm rad} + \Gamma_{nm}^{\rm ang} < c_L^2$, for $n$ and $m$
on the same shell and on different shells. In particular, there is positional
order of the radii of the different shells. However, we can distinguish between
the case that there exists oriental order between the shells, in which case the
relative angle is locked and  $\Gamma_{nm}^{\rm SS} < \frac{2 \pi}{N_v^{\rm max}}
c_L^2$. This we call the solid phase (S). When the oriental order between the
shells is destroyed we call this phase the solid of solid shells (SS).
The melted phase is charactarized by $\Gamma_{nm} = \Gamma_{nm}^{\rm rad} +
\Gamma_{nm}^{\rm ang} > c_L^2$, and in particular the positional order in the radii
of the different shells is destroyed. Within this phase it is still possible to
have a well-defined angle between vortices on the same shell, i.e.,
$\Gamma_r^{\rm ang} < c_L^2$. This we call the hexatic liquid (HL), whereas when
this order is destroyed we call it the liquid (L). These phases are
schematically indicated in Fig. \ref{phases}.
%As the fluctuations increase, the following four phases can be identified, as
%shown in Fig. \ref{phases}:
%First, a solid, %: $\Gamma^{\rm ang}_{r_n r_m} < c_L^2$, $\Gamma^{\rm
%rad}_{r_n} <
%c_L^2$, $\Gamma^{\rm rad}_{r_n r_m} < c_L^2$, $\Gamma^{\rm ang}_{r_n} < c_L^2$;
%second a solid of solid shells, for which only $\Gamma^{\rm ang}_{r r'} >
%c_L^2$. We call the shells solid, because
%both the radii of the vortices on the shell and the angles between the vortices
%on the shell display long-range order. In the same way we call the radial
%direction solid, because there is long-range order in the radii of the different
%shells. 
%,% $\Gamma^{\rm
%rad}_{r_n} < c_L^2$, $\Gamma^{\rm rad}_{r_n r_m} < c_L^2$, $\Gamma^{\rm
%ang}_{r_n} < c_L^2$; 
%Third, a solid of hexatic shells, for which  $\Gamma^{\rm ang}_{r r'} > c_L^2$
%and $\Gamma^{\rm
%rad}_{r} > c_L^2$. This phase still has long range order in the radii of the
%different shells. On the same shell the angles between the vortices are still
%well defined, but their radii have lost long range order. % $\Gamma^{\rm
%rad}_{r_n r_m} < c_L^2$, $\Gamma^{\rm
%ang}_{r_n} < c_L^2$; 
%Fourth, a liquid of hexatic shells for which only 
% $\Gamma^{\rm ang}_{r_n r_m} > c_L^2$,
%$\Gamma^{\rm rad}_{r_n} > c_L^2$, $\Gamma^{\rm rad}_{r_n r_m} > c_L^2$,
%$\Gamma^{\rm ang}_{r} < c_L^2$, i.e., only the angle between the vortices on the
%same shell is still well defined, 
%and finally the liquid.%: $\Gamma^{\rm ang}_{r_n r_m} > c_L^2$, $\Gamma^{\rm
%rad}_{r_n} >
%c_L^2$, $\Gamma^{\rm rad}_{r_n r_m} > c_L^2$, $\Gamma^{\rm ang}_{r_n} > c_L^2$.

\begin{figure}
\begin{center}
\psfrag{Omega}{$\Omega/\omega_\perp$}
\psfrag{R_Gamma/R}{ $R_{\Gamma} / R$}
\includegraphics[scale=.66]{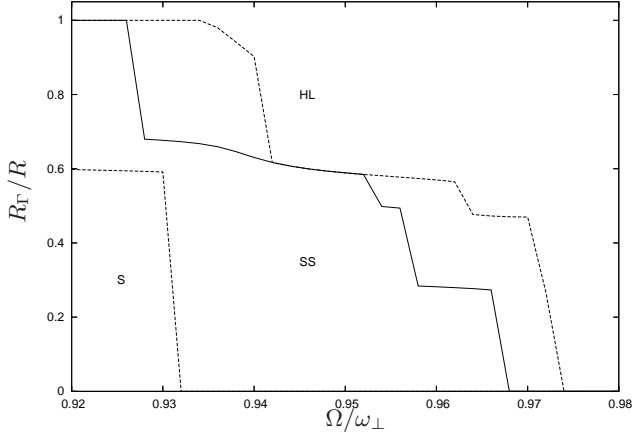}
\end{center}
\caption{Phase boundaries between the solid (S), the solid of solid shells (SS), and the hexatic liquid (HL) as a function of radial distance and 
rotation frequency for $N= 250$ and $U=10$. The solid line is for $t=0$ and the dashed line is for $t=1/100$.}
\label{hexatic}
\end{figure}

These criteria should again be applied locally because of the inhomogeneous
density, and we define the phase boundaries in analogy to the previous
definition as the radius of the innermost ring that is partly melted. 
For our parameters it turns out that the angle between neighboring vortices on 
the same ring is always well-defined, but we can identify the
transition between the solid, the solid of solid shells, and the hexatic liquid.
The result of this calculation is presented in Fig. \ref{hexatic}. In agreement
with known theory for vortex lattices, the hexatic symmetry is a very robust
phenomenon.

\section{Classical melting} \label{meltV}
The Bose-Einstein temperature for a two-dimensional noninteracting Bose gas in a
harmonic trap is given by
\begin{equation}
k_B T_c = \omega_\perp \sqrt{\zeta(2)/N},
\end{equation}
where $\zeta(x)$ is the Riemann zeta function and $\zeta(2) \simeq 1.28$. When
the temperatures are much lower than this temperature, there is no thermal cloud
present in the gas and we can easily extend our analysis to this regime. This is
experimentally relevant,  because the zero-temperature limit is difficult to
reach
\cite{Cornell}. However,
according to the Mermin-Wagner theorem, a two-dimensional crystal cannot exist
for nonzero temperatures.
For an infinite hexagonal vortex lattice this can be seen as follows. Let ${\bf
u}({\bf x}, t)$  be the displacement field of the vortex lattice. We define
\begin{eqnarray}
u_L ({\bf q}, t) &=& (q_x u_x({\bf q}, t) + q_y u_y ({\bf q}, t))/q, \\
u_T ({\bf q}, t) &=& (q_x u_y({\bf q}, t) - q_y u_x ({\bf q}, t))/q, 
\end{eqnarray} 
as the longitudinal and transverse fluctuations in the displacement field,
respectively. For long wavelengths the action for the vortex lattice is then
given by
\begin{equation}
S = \int d t \int \frac{d^2 {\bf q} }{(2 \pi)^2} \;\left(  \bar n \; u_L
\hbar \partial_t u_T - c_1 q^2
|u_L|^2 - c_{66} q^2
|u_T|^2 \right),
\end{equation}
where $\bar n$ is the average atomic density and $c_1$ and $c_{66}$ are the
compressional modulus and the shear modulus of the vortex lattice,
respectively. We have used that the vortex-vortex interaction decays faster
than $1/r^2$, since the condensate is assumed to be weakly interacting.  
From this action we read off that the dispersion of the (almost transverse) Tkachenko
mode is quadratic, i.e.,
\begin{equation}
\omega({\bf q})= 2 \sqrt{c_1 c_{66} } q^2 / \bar n .
\end{equation}
The fluctuations in the vortex positions can then be calculated
as
\begin{eqnarray}
\langle u^2 \rangle = \int \frac{d^2 {\bf q} }{(2 \pi)^2} \frac{1}{2
\bar n}\left(\sqrt{\frac{c_1}{c_{66}}}+\sqrt{\frac{c_{66}}{c_1}}\right),
\end{eqnarray}
which is a converging integral if we realize that the integration is over the
first Brillouin zone.
However, for nonzero temperature, we have to add the Bose-Einstein distribution
$1/(e^{\beta \omega(k)}-1$), which 
for long wavelengths can be approximated by $ k_B T/ \omega(k)$. 
The temperature fluctuations are therefore given by
\begin{equation}
\langle u^2 \rangle_T = k_B T  \int \frac{d^2 {\bf q} }{(2 \pi)^2} \frac{1}{2
\bar n q^2} \left(
\frac{1}{c_1} + \frac{1}{c_2}\right),
\end{equation}
which diverges logarithmically in the infrared. 
Therefore, the usual Lindemann criterion always predicts melting for infinite
isolated two-dimensional vortex lattices. For a finite system there is a natural
infrared cut-off of the divergence, but still there are large contributions
from low-lying modes. Any finite amount of tunneling, however, will turn the
system into a
three-dimensional system, where the Mermin-Wagner theorem does not apply any
more. If $k$ denotes the momentum in the $z$ direction, we can then define
\begin{eqnarray}
u_L({\bf q},k, t) &=& (q_x u_x({\bf q},k,t) + q_y u_y ({\bf q},k,t)/q, \\
u_T({\bf q},k, t) &=& (q_x u_y({\bf q},k,t) - q_y u_x ({\bf q},k,t))/q.
\end{eqnarray} 
Tunneling results in an additional term in the action that for long wavelengths
is given by
\[
\int dt \int \frac{d k}{2 \pi}  \int \frac{d^2 {\bf q} } {(2 \pi)^2}  \; J k^2
(|u_L|^2 + |u_T|^2),
\]
where $J$ is an effective hopping parameter. This still gives a quadratic
dispersion, but with an anisotropic mass, i.e.,
\begin{equation}
\omega({\bf q}, k) = 2 \sqrt{(c_1 q^2 + J k^2)(c_{66} q^2 + J k^2)}/\bar n.
\end{equation}
Writing ${\bf p} = (q_x, q_y, k)$ and using spherical coordinates the
dispersion becomes
\begin{equation}
\omega({\bf p}) = 2 p^2 \sqrt{(c_1 \sin^2 \theta + J \cos^2 \theta) (c_{66} \sin^2
\theta + J \cos^2 \theta)} /\bar n.
\end{equation}
The fluctuations can therefore be calculated as
\begin{eqnarray}
&& \hspace{-.6cm} \langle u^2 \rangle_T = k_B T  \int \frac{d  p d \phi }{(2 \pi)^2}  \frac{p^2
\sin\theta}{ 2 \bar n p^2} 
 \times  \\
 && \hspace{-.2cm}
\left\{ \frac{1}{\sqrt{(c_1 \sin^2 \theta + J \cos^2 \theta) (c_{66} \sin^2 \theta
+ J \cos^2 \theta)}} \times 
\right. \nonumber \\
&& \left. 
\left(
\sqrt{\frac{c_1 \sin^2 \theta + J \cos^2 \theta}{c_{66} \sin^2 \theta  + J \cos^2
\theta}} +\sqrt{\frac{c_{66} \sin^2 \theta + J \cos^2 \theta}{c_{1} \sin^2 \theta + J
\cos^2 \theta}}\right) \right\}. \nonumber
\end{eqnarray}
which clearly converges in the infrared and the fluctuations remain finite. 
For a system of coupled pancake Bose-Einstein condensates we can therefore still
use the usual Lindemann criterion.

\begin{figure}
\begin{center}
\psfrag{Omega}{$\Omega/\omega_\perp$}
\psfrag{R_cr}{ $R_{\rm cr} / R$}
\psfrag{omega}{ $\omega(\Omega)$}
\includegraphics[scale=.66]{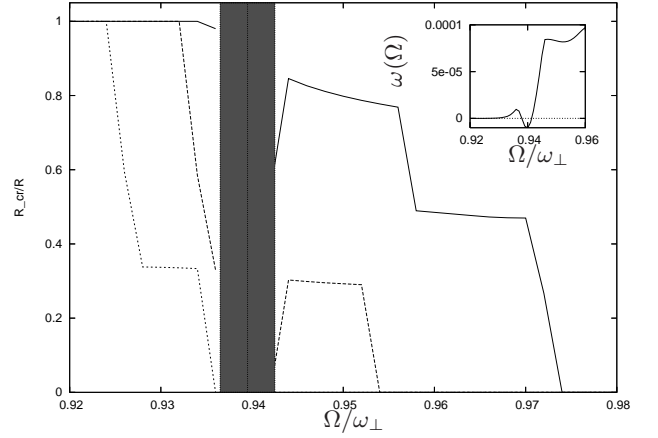}
\end{center}
\caption{Crystal radius $R_{\rm cr}$ normalized to the condensate radius $R$ as
a function
of the rotation frequency for $N= 250$, $U=10$, $t = 1/20$, and $T=T_c/40$
(dotted line), $T=T_c/50$ (dashed line), and
$T=T_c/100$ (solid line).  In the shaded region there is a dynamical instability
towards elleptic shape deformation. In the inset the frequency of the unstable
mode is plotted. }
\label{melt_temp}
\end{figure}

\begin{figure}
\begin{center}
\psfrag{T_cross}{$T_c/T_{\rm cross}$}
\psfrag{Omega}{$\Omega/\omega_\perp$}
\includegraphics[scale=.66]{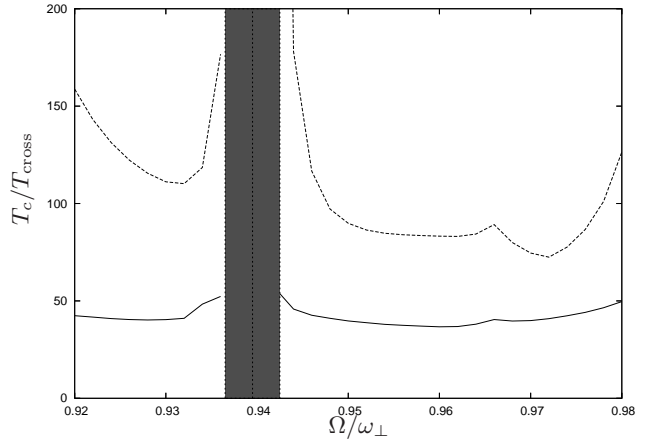}
\end{center}
\caption{Inverse of the crossover temperature $T_{\rm cross}$ 
a function 
of $\Omega/\omega_\perp$ for $N= 250$ and $U=10$. Lines are for $t = 1/20$
(solid line) and $t=0$ (dashed line).
As in Fig. \ref{melt_temp}, the shaded region  is excluded because of the
presence of a dynamical instability.}
\label{crosstemp}
\end{figure}

\begin{figure}
\begin{center}
\psfrag{T_c/T}{$T_c/T$}
\psfrag{R_cr}{ $R_{\rm cr} / R$}
\includegraphics[scale=.66]{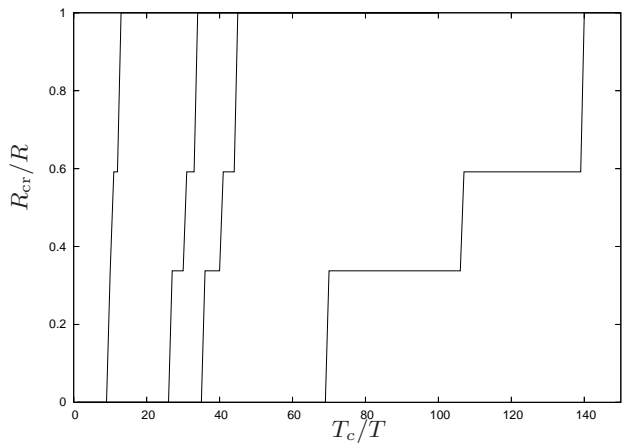}
\end{center}
\caption{Crystal radius $R_{\rm cr}$ normalized to the condensate radius $R$ as
a function 
of $T_c/T$ for $N= 250$, $U=10$, and $\Omega/\omega_\perp=0.93$. From left to
right the lines have hopping parameter  $t = 1$, $t=0.1$,  $t = 0.05$, and $t =
0.01$, respectively.}
\label{tempmelt}
\end{figure}

In Fig. \ref{melt_temp} the vortex-crystal radius is plotted as a function of
the rotation frequency for a fixed and strong tunneling and various temperatures. 
In the regime $\Omega/\omega_\perp \in \lbrack .938, .941 \rbrack$ there is a
dynamical instability towards elliptic shape deformation. This is indicated by a
shaded region in the plotted figures. We did not calculate the fluctuations in
this regime. This is related to the elliptic shape deformation that occurs
before a single vortex enters the condensate, and that has been investigated
before theoretically \cite{Dalfovo01, Recati01, Sinha01, Kramer02} and has also
been observed \cite{Madison01}. Because the unstable mode crosses zero, this
causes huge fluctuations in the neighborhood of this instability.  
In Fig. \ref{crosstemp} we calculate the temperature for which quantum
fluctuations of the vortex crystal are equal to the 
temperature fluctuations, which defines the crossover temperature between
quantum and classical melting. The temperature should be chosen much lower to
observe the effects of quantum melting. 
In Fig. \ref{tempmelt} the vortex-crystal radius is plotted as a function of
temperature for a fixed rotation frequency.

\begin{figure}
\begin{center}
\psfrag{Omega}{$\Omega/\omega_\perp$}
\psfrag{R_Gamma/R}{ $R_{\Gamma} / R$}
\includegraphics[scale=.66]{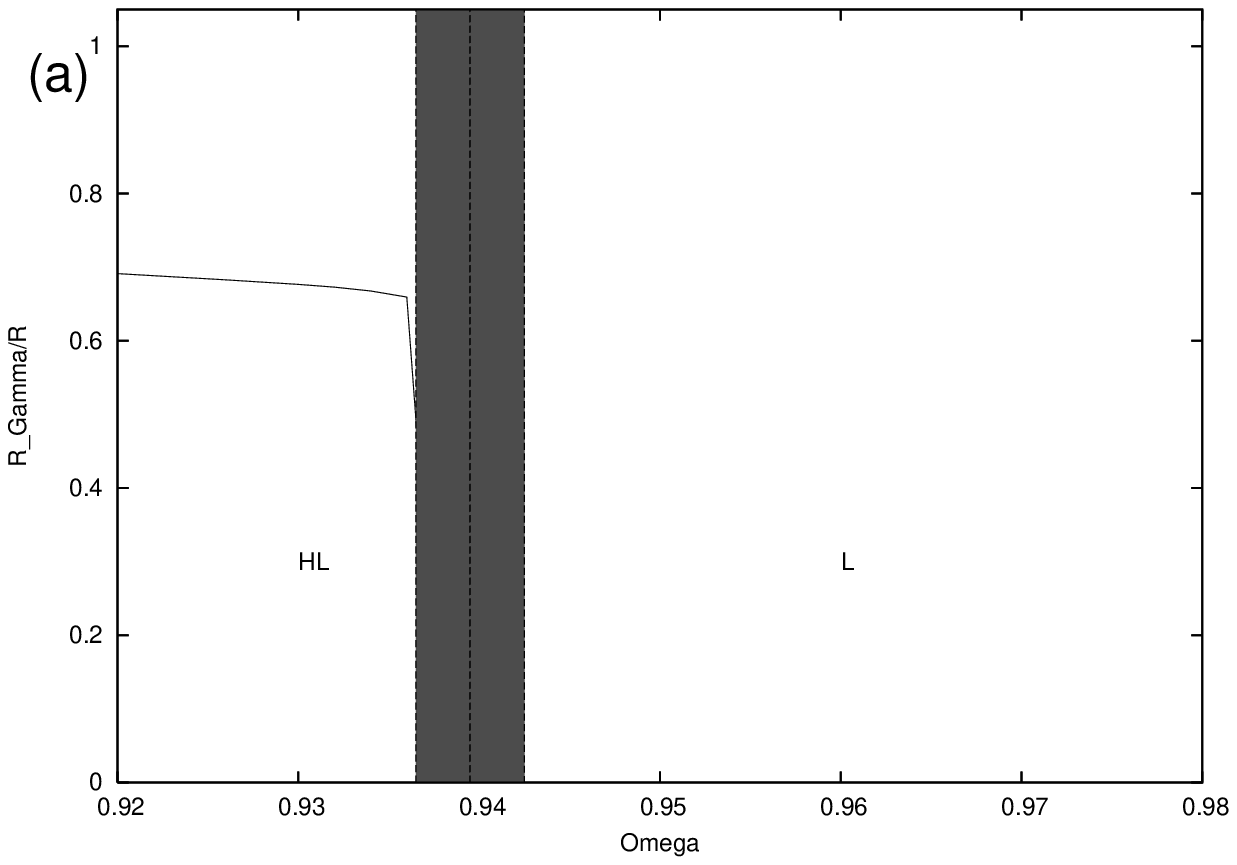}
\includegraphics[scale=.66]{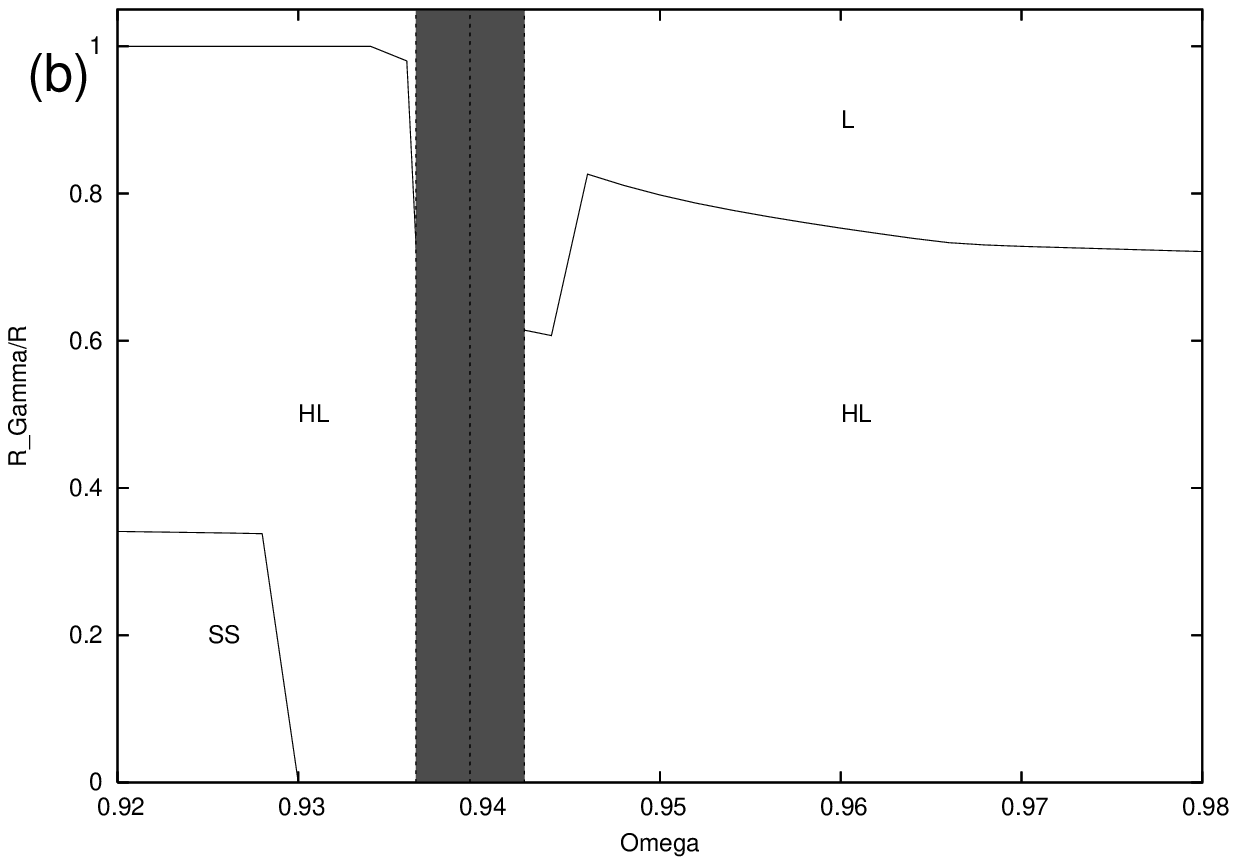}
\includegraphics[scale=.66]{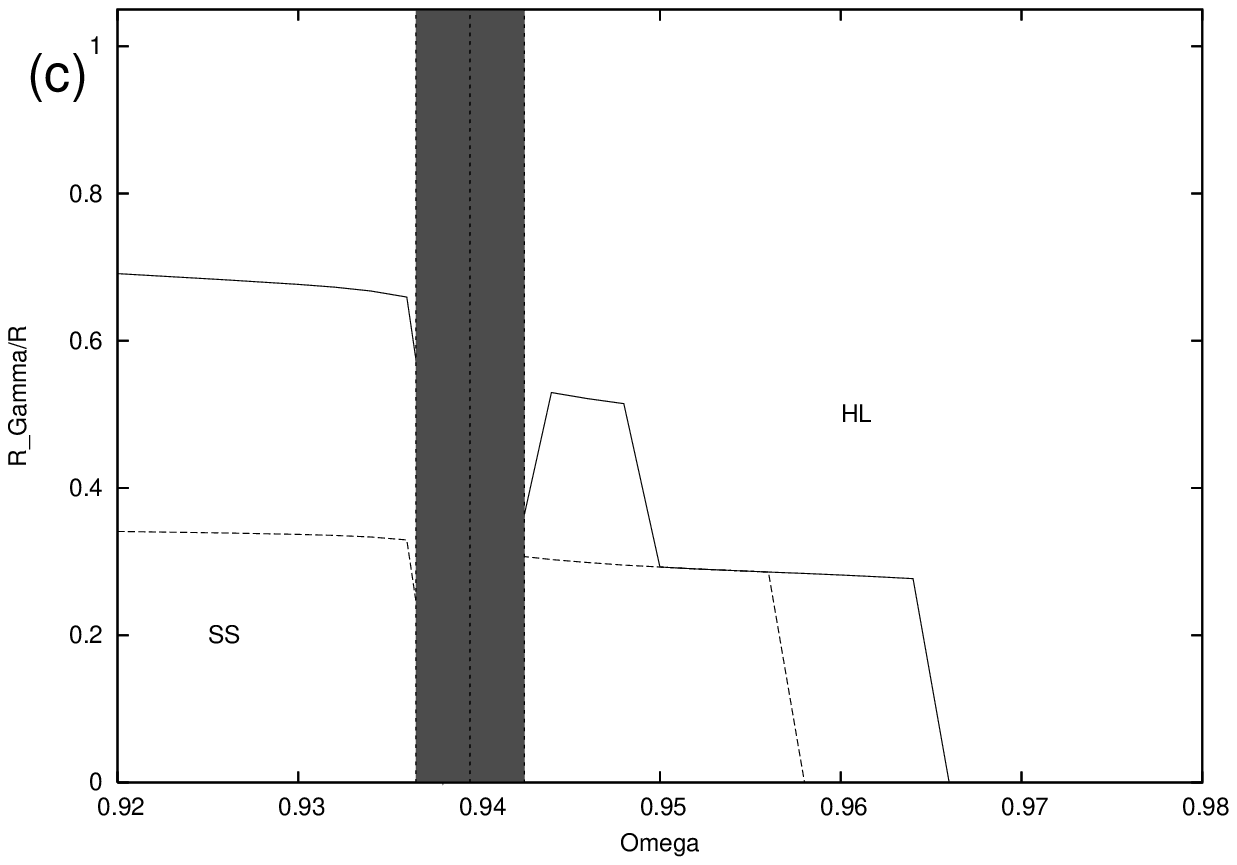}
\end{center}
\caption{Phase boundaries between the solid of solid shells (SS), hexatic liquid (HL) and liquid (L), as a function of radial distance and 
rotation frequency for $N= 250$, $U=10$, and $t=0$ (dashed line) and $t=1/20$ (solid lines). 
The figures are for a) $T=T_c/20$,  b) $T=T_c/35$, and c) $T=T_c/100$.}
\label{hexatictemp}
\end{figure}
When tunneling between sites is suppressed, the correlation between neighboring
vortices
\begin{equation}
\frac{\langle ({\bf u}_{in} - {\bf u}_{im})^2\rangle}{2 \Delta_{mn}^2},
\end{equation}
where $\Delta_{mn}$ is the distance between the vortex $n$ and $m$ at site $i$,
has been proposed as an appropriate order parameter,
with unchanged Lindemann parameter \cite{Zahn99, Baym04, Kierfeld04, Dietel06}.
In this order parameter the diverging small
momentum contributions are subtracted. Nevertheless, it turns out that we have
to go to low temperatures
to see any crystalline order. The same phases as in the case of zero temperature can be distinguished.
The result of this calculation is plotted in Fig. \ref{hexatictemp}.
Note that a true solid phase of the vortex lattice is not observable in these phase diagrams, not even for the lowest temperature of $T=T_c/100$.

%As can be seen in this figure there seems to appear a little area in the phase
%diagram where the angular order 
%within the ring is already destroyed, but not the radial order between the
%rings. This phase may be called a
%solid of liquid rings (SL).

%We also computed the correlation with the
%central vortex:
%\begin{equation}
%\frac{\langle ({\bf u}_{in} - {\bf u}_{i0})^2\rangle}{\Delta_{n0}^2},
%\end{equation}
%and 
%\begin{equation}
%\frac{\langle {\bf u}^2_{in} \rangle - \langle {\bf u}_{in} {\bf
%u}_{i0}\rangle}{\Delta_{n0}^2},
%\end{equation}
%which have the same effect. This last order parameter has the advantage that it
%does not give the average of  
%the fluctuations but only subtracts the correlation with the central vortex.
%The
%disadvantage is, that this order parameter can become negative, but in practice
%the central
%vortex always has the smallest fluctuations, such that this is not a problem.
%Because of the inhomogeneous system we are dealing with, the correlations will
%depend on the distance to the origin, but also on the angle. 

\section{Anharmonic radial confinement} \label{meltVI}
\begin{figure}
\begin{center}
\includegraphics[scale=.75]{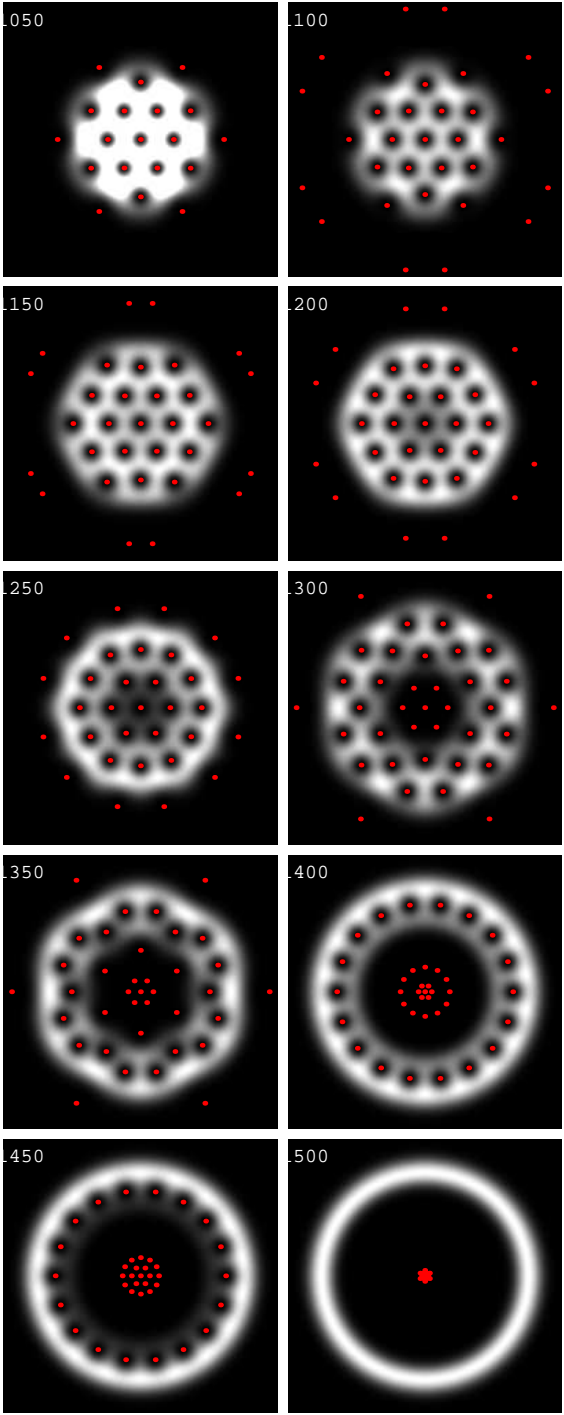}
\end{center}
\caption{(Color online) Classical vortex lattice and density profile 
for rotation frequencies $\Omega/\omega_\perp=1.05, 1.1, \ldots, 1.5$ in the
presence of a quartic potential. Parameters are chosen as $U=10$ and
$\lambda=0.01$.
White means high density, black low density. The vortex positions are indicated
by a dot, such that
also the vortices outside the condensate are visible.}
\label{quartic_pic}
\end{figure}
\begin{figure}
\begin{center}
\psfrag{Omega}{$\Omega/\omega_\perp$}
\psfrag{R}{ $R$}
\includegraphics[scale=.66]{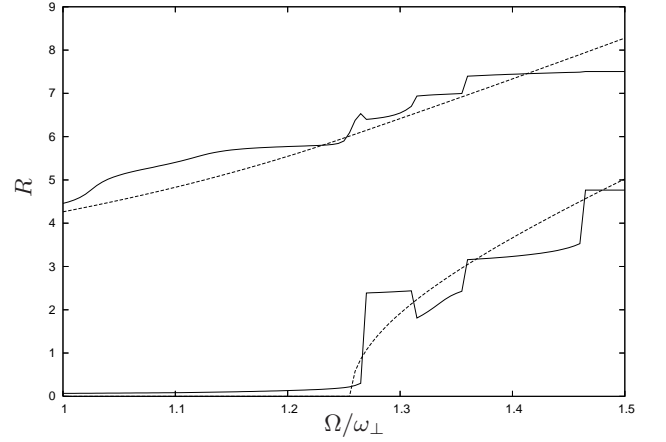}
\end{center}
\caption{Inner and outer radius of the condensate as a function 
of the rotation frequency for $N= 250$, $U=10$, $\lambda=0.01$, and $t=0$.
The solid line is the result of a calculation using the lowest Landau
approximation and with 37 vortices taken into account. The dashed line is the
result of the approximation that the density profile is described by a quartic
polynomial.}
\label{radii}
\end{figure}
The inclusion of a quartic potential, in addition to the usual harmonic
potential, has attracted interest for several reasons.
The quartic potential plays an important role in the understanding of vortex
nucleation \cite{Fetter01, Gosh04}. For our purposes, however, it is more
important that in this way rotation frequencies larger than the trap frequency
can be applied. This causes the density to be lower in the center of the trap,
and eventually gives rise to the formation of giant vortices, i.e., multiple
quantized vortices, in the center of the trap \cite{Lundh02, Kasamatsu02,
Kavoulakis03}.
Experiments in this setup were performed and it was observed that the vortex
lattice became disordered, but no giant vortex was observed \cite{Bretin04}.
Giant vortex formation was observed by artificially removing the inner part of a
fast rotating condensate by means of a laser beam \cite{Engels02}.  
Shape oscillations of a vortex lattice in an anharmonic potential
were also studied experimentally \cite{Stock04}.
Theoretically, the phase diagram of this system was studied intensively to
identify the parameter space where giant vortex formation can take place
\cite{Aftalion04, Jackson04, Fetter04}. 
Other aspects that are studied are the dynamics of forming the giant vortex
\cite{Tsubota03, Fu05}, oscillations of the vortex lattice \cite{Cozzini05},
aspects on observation \cite{Danaila05, Cozzini06}, and stability of quantum
fluctuations \cite{Kim05}. 

It is straightforward to extend our analysis to this case. We first of all add
the quartic potential
\begin{equation}
V_4(r) = \lambda \frac{r^4}{\ell_\perp^4},
\end{equation}
where $\ell_\perp$ is the harmonic length associated with the radial trapping.
We again use lowest Landau level wavefunctions and make the ansatz that there is
one vortex in the center and that the rest of the vortices order themselves in
rings of multiples of six. The resulting density profiles for 37 vortices are
plotted in Fig. \ref{quartic_pic} as a function of the rotation frequency.
We compare these density profiles with the Thomas-Fermi-like density profiles of
the form 
\begin{equation}
n(r) \propto \left(\frac{r^2}{R_1^2} \pm 1 \right)\left(\frac{r^2}{R_2^2} -
1\right), 
\end{equation}
that come from minimizing the on-site energy functional without paying attention
to the lowest Landau level constraint.
When the minus sign is chosen there appears a hole in the density, and the
condensate shape is annular with inner radius $R_1$ and outer radius $R_2$. We
compare the values coming from this ansatz with the inner and outer radius
coming from the lowest Landau level densities in Fig. \ref{radii}.
They agree quite well. 

We also calculate the quantum fluctuations of the vortices. This is only
possible in a limited regime, where the ansatz of a central vortex surrounded by
rings of six vortices is dynamically stable. In this regime, the inner part of
the vortex crystal is melted, although the particle density is nonzero there. We
define the liquid radius $R_{\rm L}$   as the radius of the innermost vortex
that is part of the vortex crystal.
In the region $\Omega/\omega_\perp \in \lbrack 1.155, 1.25 \rbrack$ the ansatz
is stable against fluctuations. We find that the liquid radius in this region is
almost constant and given by $R_{\rm L}/R \simeq .58$. 
There is a second radius which separates the vortex crystal from the vortex
liquid that is at the outside of the condensate, which corresponds to the
crystal radius $R_{\rm cr}$ defined before. In this region this radius is always
equal to the condensate radius or $R_{\rm cr}/R = 1.0$.

\section{Conclusions and outlook} \label{meltVII}
In this article we derived the theory of vortex fluctuations in a
one-dimensional optical lattice. We showed that in this configuration the modes
of the vortex lattice get a dispersion in the axial direction. In particular we
discussed the Bloch bands of the Tkachenko modes. Based on the Lindemann
criterion we studied the melting of the vortex lattice. Because of the
inhomogeneous density the melting occurs from the outside inwards. Comparison
with a local density-approximation yields important finite-size corrections,
because the local-density theory does not take into account the discrete nature
of the vortices and overestimates the rotation frequency that is needed for
total melting of the vortex lattice. Tunneling between the sites of the optical
lattices decreases the fluctuations and causes freezing of the vortex lattice.
Temperatures on the other hand increases the fluctuations considerably. The
crossover temperature from classical to thermal melting is very low, which makes
it an experimentally difficult problem to see the effects of quantum
fluctuations. 
By looking into the correlations between the vortices, several phases can be
distinguished, where part of the order is destroyed by quantum fluctuations.

In the one-dimensional optical lattice there is a clear experimental signal for
both the fluctuations in the vortex position and the liquid. The fluctuations
can be measured in analogy with 
the situation of a single vortex in an optical lattice. By imaging in axial
direction one will see the vortex cores
distributed themselves in a gaussian distribution around their equilibrium
position, from which the size of the fluctuations can be extracted
\cite{Martikainen04}.  
In the liquid the vortices are no longer individually visible, which is a clear
distinction from the vortex crystal. 
An interesting question is whether the liquid will completely restore the
rotation symmetry that is broken by the presence of the vortex lattice, or the
liquid is partly pinned because of the interaction between the vortex liquid and
the vortex crystal. 

It remains a challenging problem
to describe the coexisting crystal-liquid. This will allow to decide on the
occurrence of melting based
on energy considerations and thus shed more light on
the accuracy of the application of the Lindemann criterion in this
inhomogeneous situation. This also applies to the phases where partial 
melting takes place. 
% whereas for other correlation function still
% long-range order can be found. 
It is not only important to find a good
description of these phases themselves,
but also the predicted phase coexistence between them raises interesting new
questions.  

\section*{Acknowledgments}
We thank Nigel Cooper, Masud Haque, Jani Martikainen, Hagen Kleinert, J\"urgen
Dietel, and Alexander Fetter for useful
discussions. 
This work is supported by the Stichting voor
Fundamenteel Onderzoek der Materie (FOM) and the Nederlandse
Organisatie voor Wetenschappelijk Onderzoek (NWO).

\end{document}